# A Virtual Element Method for 2D linear elastic fracture analysis

---

## Contents

*Email addresses:* zhuang@ikm.uni-hannover.de (Xiaoying Zhuang ), ngx.hung@hutech.edu.vn (Hung Nguyen-Xuan )

# A Virtual Element Method for 2D linear elastic fracture analysis

Vien Minh Nguyen-Thanh[a], Xiaoying Zhuang[a,*], Hung Nguyen-Xuan[b,d,*], Timon Rabczuk[c], Peter Wriggers[a]

[a]*Institute of Continuum Mechanics, Leibniz Universität Hannover, Appelstraße 11, 30167 Hannover, Germany*
[b]*Center for Interdisciplinary Research in Technology, Ho Chi Minh City University of Technology (Hutech), Ho Chi Minh City, Vietnam*
[c]*Institute of Structural Mechanics, Bauhaus-University Weimar, Marienstraße 15, 99423 Weimar, Germany*
[d]*Department of Architectural Engineering, Sejong University, 209 Neungdong-ro, Gwangjin-gu, Seoul 05006, Republic of Korea*

**Abstract**

This paper presents the Virtual Element Method (VEM) for the modeling of crack propagation in 2D within the context of linear elastic fracture mechanics (LEFM). By exploiting the advantage of mesh flexibility in the VEM, we establish an adaptive mesh refinement strategy based on the superconvergent patch recovery for triangular, quadrilateral as well as for arbitrary polygonal meshes. For the local stiffness matrix in VEM, we adopt a stabilization term which is stable for both isotropic scaling and ratio. Stress intensity factors (SIFs) of a polygonal mesh are discussed and solved by using the interaction domain integral. The present VEM formulations are finally tested and validated by studying its convergence rate for both continuous and discontinuous problems, and are compared with the optimal convergence rate in the conventional Finite Element Method (FEM). Furthermore, the adaptive mesh refinement strategies used to effectively predict the crack growth with the existence of hanging nodes in nonconforming elements are examined.

*Corresponding author
*Email addresses:* zhuang@ikm.uni-hannover.de (Xiaoying Zhuang ), ngx.hung@hutech.edu.vn (Hung Nguyen-Xuan )

## 1 Introduction

The VEM, introduced in [1], can be considered as a generalization of FEM which utilizes general meshes in complex geometries. In fact, the ancestor of VEM is the mimetic finite difference method [2] but owing to restrictions to low order approximations for its applications, VEM was established to limit its forefather's disadvantages and to bring its key advantages into numerical applications. Although VEM has been recently introduced, it has attracted a great number of attentions from the numerical scientific community and has numerous results from the general applications such as linear elasticity problems in 2D and 3D [3, 4, 5], inelastic problems [6, 7], compressible and incompressible nonlinear elasticity [8], electromagnetic problems [9], geomechanical simulations of reservoir [10], contact problem [11], topology optimization [12] to the specific issues of VEM, e.g. serendipity [13], virtual and smoothed finite elements [14], conforming and nonconforming VEM [15], discontinuous Galerkin VEM [16], high-order VEM [17], stability analysis [18], hourglass stabilization [19], mixed VEM [20], $\mathcal{L}^2$ projection in VEM [21], and the studies related to VEM have been broadening very fast in recent years. Though a background of VEM will be rigorously discussed in the next sections, VEM can be recapitulated in three key aspects [22]. Firstly, the virtual element space consists of both polynomials of degree up to $k$ and some non-polynomial functions. Secondly, one part of the local stiffness matrix called consistency term in which one of two entries is a polynomial of degree up to $k$ and this term is exactly computable with the prescribed degrees of freedom. Thirdly, the other part of the local stiffness matrix known as the stability term in which both entries are non-polynomial functions is approximated by choosing a "nice" portion so that the stiffness bilinear form is approximated to the right order of magnitude.

Although the VEM in the present work aims to modeling discontinuity in LEFM, there are relevant studies in the field of discrete fracture network reported by Benedetto [23, 24]. Also, Fumagalli and Keilegavlen [25] have circumvented the complex requirements on grid construction and have simplified the flow discretization with the aid of VEM to compute the pressure field and Darcy velocity in the simulation of underground fluid flow in fractured media. Another research related to our current work is to study the convergence of VEM for mechanics problems with the presence of corner singularities addressed by L. B. da Veiga et al. [26]. Herein, a mathematical finding of a priori convergence theory to the $p-$version and $hp-$version of VEM for non-smooth solutions having typical corner singularities was conducted.

Due to the beneficial and intriguing mesh flexibility, VEM is suitable for crack analysis and adaptive mesh refinement. Although some polygonal or polyhedral element methods, e.g. polygonal finite element method (PFEM) [27], polygon scaled boundary finite element method (PSBFEM) [28, 29], or extended FEM on polygonal and quadtree meshes (XPFEM) [30], are available and show their capability in crack modeling, VEM seems to be more versatile in the aspect of both calculation and refinement than the others because the explicit formulation of integral shape functions is ingeniously invisible and that makes VEM "virtual". In addition, a projection is employed as a substitution by making use of known vertices having predefined properties. Thus, the node number within an element is arbitrary.

This paper focuses on VEM's formulation and implementation for linear crack problems. Owing to the flexibility of discretization regardless of conforming or nonconforming, convex or concave elements, we carry out a posteriori error estimation to perform a $h$-adaptive mesh refinement. We will take the newest vertex bisection (NVB) [31] for triangular elements, middle point (midPoint) [32, 33] for polygonal elements and, recently, polyTree [34] for voronoi cells as the refinement strategies in the present work. In this context, the SPR method addressed by Zienkiewicz & Zhu [35, 36] is modified for arbitrary element shapes to efficiently approximate the analytical solutions. An energy method, furthermore, is used to predict the stress state in the vicinity of a crack tip imposed by remote loads. Among the virtual crack extension [37, 38], $J$-integral [39], [40] and others, we adopt and alter the $J$-integral to obtain fracture parameters and then integrate it in the VEM. Therefore, the integration can be computed over polygonal elements.

The outline of the paper is the following. Section 2 gives the brief introduction of governing equations in linear elastic fracture mechanics. Here, the basic concept of VEM with traditional projection operator, its formulation, and the calculation of both stiffness matrix and the right hand side are also presented. In Section 3, we show the computation of stress intensity factors in VEM. Afterwards, we present the adaptive schemes based on the superconvergent path recovery for polygonal elements and posteriori error estimates. From that result, the crack propagation criteria will be introduced. The numerical results are then presented to demonstrate the meshing versatility and its convergence for both continuous and discontinuous problems in Section 4. The last section will close the paper with concluding remarks.

## 2 Virtual Element Method for 2D elasticity

### *2.1 Governing equations*

Let us consider a domain $\Omega$ in two-dimensional space with an internal crack, denoted by $\Gamma_c$, as shown in Figure 1. Herein, $\bar{\boldsymbol{t}}_\sigma$ and $\bar{\boldsymbol{t}}_c$ are the prescribed tractions applied on the external and on the crack boundaries, respectively; $\boldsymbol{n}_\sigma$ and $\boldsymbol{n}_c$ are the unit normal vectors outward to each of corresponding surfaces. The boundaries, moreover, are assumed to satisfy $\Gamma_\sigma \cup \Gamma_c \cup \Gamma_u = \Gamma$, $\Gamma_\sigma \cap \Gamma_c \cap \Gamma_u = \emptyset$. All of the work are studied under the small deformation within static condition.

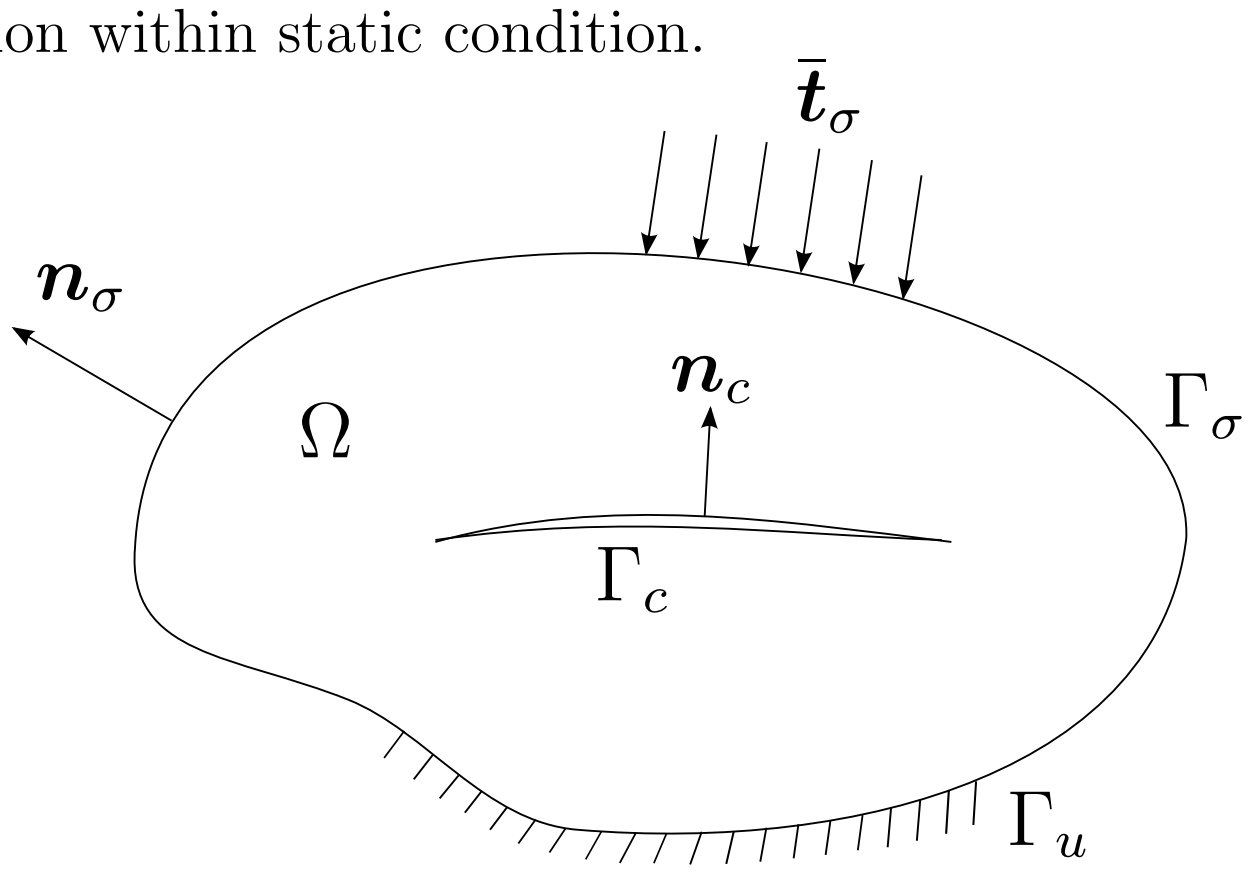

Figure 1: A two-dimensional domain $\Omega$ with a middle crack interface $\Gamma_c$.

The governing equation of the system is written in form of

$$\partial_j \sigma_{ij} + b_i = 0 \quad \forall\, i, j = 1, 2, \tag{1}$$

where $\partial$ is the differential operator with respect to Cartesian coordinate, $\boldsymbol{\sigma}$ is the Cauchy stress tensor, and $\boldsymbol{b}$ is the body load vector. The stress $\boldsymbol{\sigma}$ is related to the strain $\boldsymbol{\varepsilon}$ by the constitutive law of isotropic elastic materials

$$\sigma_{ij} = C_{ijkl}\, \varepsilon_{kl}, \tag{2}$$

where four order tensor $\mathbb{C}$ is the material tensor. The relation between strain and displacement is known through the compatibility relation

$$\varepsilon_{ij} = \frac{1}{2}\left(\partial_j u_i + \partial_i u_j\right). \tag{3}$$

In addition, the boundary conditions are depicted on the Neumann boundary ($\Gamma_\sigma$), Dirichlet boundary ($\Gamma_u$), and the crack interfaces ($\Gamma_c$)

$$\sigma_{ij} n_{\sigma j} = \bar{t}_{\sigma i} \quad \forall \mathbf{x} \in \Gamma_\sigma, \tag{4}$$

$$\sigma_{ij} n_{cj} = \bar{t}_{ci} \quad \forall \mathbf{x} \in \Gamma_c, \tag{5}$$

$$u_i = \bar{u}_i \quad \forall \mathbf{x} \in \Gamma_u. \tag{6}$$

Multiplying equations (1), (4), (5) by arbitrary displacement test functions $\delta \boldsymbol{u}$ and taking the integral over the entire domain, we obtain

$$\int_\Omega \delta u_i \left(\partial_j \sigma_{ij} + b_i\right) d\Omega + \int_{\Gamma_c} \delta u_i [\![ \sigma_{ij} n_{cj} - \bar{t}_{ci} ]\!] \, d\Gamma_c + \int_{\Gamma_\sigma} \delta u_i \left(\sigma_{ij} n_{\sigma j} - \bar{t}_{\sigma i}\right) d\Gamma_\sigma = 0. \tag{7}$$

Applying the Gaussian theorem with the traction free assumed on the crack interfaces, this equation can be rewritten as follows

$$\int_\Omega \delta \varepsilon_{ij} \sigma_{ij} \, d\Omega = \int_{\Gamma_\sigma} \delta u_i \bar{t}_{\sigma i} \, d\Gamma_\sigma + \int_\Omega \delta u_i b_i \, d\Omega. \tag{8}$$

This equation is known as the weak form for the governing equation which can be solved approximately on an appropriate finite-dimensional subspace.

### *2.2 Virtual Element Method*

In this section, we introduce the virtual discretization of the problem (8). In what follows, we will keep the notations as presented in [6]. Let a mesh $\Omega_h$ be a decomposition of the domain $\Omega$ into finite number of non-overlapping polygons $K$ in $\mathbb{R}^2$. We denote by $E \in \partial K$ the generic edge. Also, let $\boldsymbol{x}_K$ be the centroid, $h_K$ be the diameter, and $|K|$ be the area of the element $K$.

A local virtual space is defined, that is

$$\boldsymbol{V}_{h,K} := \{\boldsymbol{v} \in [H^1(K) \cap C^0(K)]^2 : \Delta \boldsymbol{v} \in \boldsymbol{\mathcal{P}}_{k-2}(K) \text{ in } K,\ \boldsymbol{v}_{|E} \in \boldsymbol{\mathcal{P}}_k(E)\ \ \forall E \in \partial K\}, \tag{9}$$

where $\boldsymbol{\mathcal{P}}_k(E)$ is a set of polynomials of degree up to $k$ on $E$ and $\Delta$ denotes the component-wise Laplace operator. More specifically, $\boldsymbol{V}_{h,K}$ contains all polynomials of degree up to $k$ and other functions on the entire local domain but only polynomials are visible on its edges. A function $\boldsymbol{v} \in \boldsymbol{V}_{h,K}$ is defined by the following properties:

- $\boldsymbol{v}$ is a polynomial of degree $k$ on each edge $E$, i.e. $\boldsymbol{v}_{|E} \in \boldsymbol{\mathcal{P}}_k(E)$.

- $\boldsymbol{v}$ on $\partial K$ is globally continuous, i.e. $\boldsymbol{v}_{|\partial K} \in \boldsymbol{C}^0(\partial K)$.
- $\Delta \boldsymbol{v}$ is a polynomial of degree $k-2$ in $K$, i.e. $\Delta \boldsymbol{v} \in \boldsymbol{\mathcal{P}}_{k-2}(K)$.

Now the global virtual element space can be defined based on the local space in the following

$$\boldsymbol{V}_h := \{\boldsymbol{v} \in \boldsymbol{V} := [H_0^1(K)]^2 : \boldsymbol{v}_{|K} \in \boldsymbol{V}_{h,K} \;\; \forall K \in \Omega_h\}. \tag{10}$$

Thus, the dimension of $V_h$ is

$$N := N_V + N_E(k-1) + N_P \frac{k(k-1)}{2}, \tag{11}$$

where $N_V$, $N_E$, and $N_P$ are the total number of internal vertices, internal edges, and elements, respectively, in $\Omega_h$.

In our study, we merely consider the polynomials of degree up to order $k = 1$. Thus, the space $\boldsymbol{V}_{h,K}$ is a space of harmonic functions which are piecewise linear and continuous on the boundary of the element. Such space is "virtual" in the sense which is well defined but not known explicitly inside the element.

Next, one the most prominent feature of VEM is the projection operator. Here, it is defined in the following

$$\Pi_K^{\nabla} : \boldsymbol{V}_{h,K} \to \boldsymbol{\mathcal{P}}_1(K). \tag{12}$$

Provided that there is no confusion, we can use $\Pi$ instead of $\Pi_K^{\nabla}$. With the defined operator we have the following orthogonality condition as the energy projection. It reads

$$a^K(\boldsymbol{p}, \boldsymbol{v} - \Pi \boldsymbol{v}) = 0. \tag{13}$$

Consequently, we also have a projection operator onto constants denoted by $\mathsf{P}_0$

$$\mathsf{P}_0(\Pi \boldsymbol{v} - \boldsymbol{v}) = 0. \tag{14}$$

Here, $\mathsf{P}_0$ in linear case can be chosen as follows

$$\mathsf{P}_0(\boldsymbol{v}) = \frac{1}{n_V} \sum_{i=1}^{n_V} \boldsymbol{v}(\boldsymbol{X}_i), \tag{15}$$

where $n_V$ is the number of vertices of the polygon $K$. With the projection operator (12), analogously to FEM, the approximation of the projection of an arbitrary function, i.e. displacement field, in $\boldsymbol{V}_h$ can be written as follows

$$\Pi \boldsymbol{u} \approx \sum_{i=1}^{n_V} \Pi \boldsymbol{\varphi}^i \hat{\boldsymbol{u}}^i = \Pi \boldsymbol{\varphi} \cdot \hat{\boldsymbol{u}}, \tag{16}$$

where $\Pi\boldsymbol{\varphi}$ is the projection of basis functions, $\hat{\boldsymbol{u}}$ is the nodal displacement vector. In the matrix form, they are presented in the following

$$\Pi \boldsymbol{\varphi} = \begin{bmatrix} \boldsymbol{\Pi\varphi}^1 & \boldsymbol{\Pi\varphi}^2 & \dots & \boldsymbol{\Pi\varphi}^{n_V} \end{bmatrix}, \tag{17}$$

$$\hat{\boldsymbol{u}} = \begin{bmatrix} \hat{u}_1^1 & \hat{u}_2^1 & \hat{u}_1^2 & \hat{u}_2^2 & \dots & \hat{u}_1^{n_V} & \hat{u}_2^{n_V} \end{bmatrix}^T. \tag{18}$$

The basis functions $\varphi^i \in V(K)$ are defined as usual as the canonical basis functions

$$\varphi^i(\boldsymbol{X}_j) = \delta_{ij}, \qquad i, j = 1, 2, ..., n_V. \tag{19}$$

Because $\Pi\boldsymbol{\varphi}$ belongs to polynomial space, we can express it as follows

$$\Pi \boldsymbol{\varphi} = \mathbf{M} \cdot \boldsymbol{\Pi}_* \tag{20}$$

and its components are

$$\Pi \boldsymbol{\varphi}^i = \mathbf{M} \cdot \boldsymbol{\pi}^i = \sum_{\alpha=1}^{n_k} \boldsymbol{m}^\alpha \boldsymbol{\pi}^{\alpha i}, \tag{21}$$

where $\boldsymbol{\pi}^i$ are the coefficients of the projection of $\boldsymbol{\varphi}^i$, and $\mathbf{M}$ is a matrix whose components $\boldsymbol{m}^\alpha$ are basic functions of the linear space ($k = 1$). Please note that the dimension of $\boldsymbol{\mathcal{P}}_1(K)$ space in equation (9) is 3, so $n_k$ is 6. Therefore, $\mathbf{M}$ is chosen as follows

$$\mathbf{M} = \begin{bmatrix} 1 & 0 & -y + y_K & y - y_K & x - x_K & x - x_K \\ 0 & 1 & x - x_K & x - x_K & -y + y_K & y - y_K \end{bmatrix}. \tag{22}$$

For any linear function $\boldsymbol{p} \in \boldsymbol{\mathcal{P}}_1(K)$ and $\boldsymbol{v} \in \boldsymbol{V}_{h,K}$, applying the Gaussian theorem to transfer the domain integral to the boundary integral and because the divergence of stress, which is a function of linear functions, is zero, we

obtain

$$
\begin{aligned}
a^K(\boldsymbol{p}, \boldsymbol{v}) &= \int_{\Omega_K} \boldsymbol{\varepsilon}(\boldsymbol{v}) : \boldsymbol{\sigma}(\boldsymbol{p})\, d\Omega \\
&= \int_{\Omega_k} div\left(\boldsymbol{v} \cdot \boldsymbol{\sigma}(\boldsymbol{p})\right) d\Omega - \int_{\Omega_K} \boldsymbol{v} \cdot div(\boldsymbol{\sigma}(\boldsymbol{p}))\, d\Omega \\
&= \int_{\Gamma_K} \boldsymbol{v} \cdot \boldsymbol{\sigma}(\boldsymbol{p}) \cdot \boldsymbol{n}\, d\Gamma. \qquad (23)
\end{aligned}
$$

The rest boundary integral is probably computed by means of Gauss Lobatto.

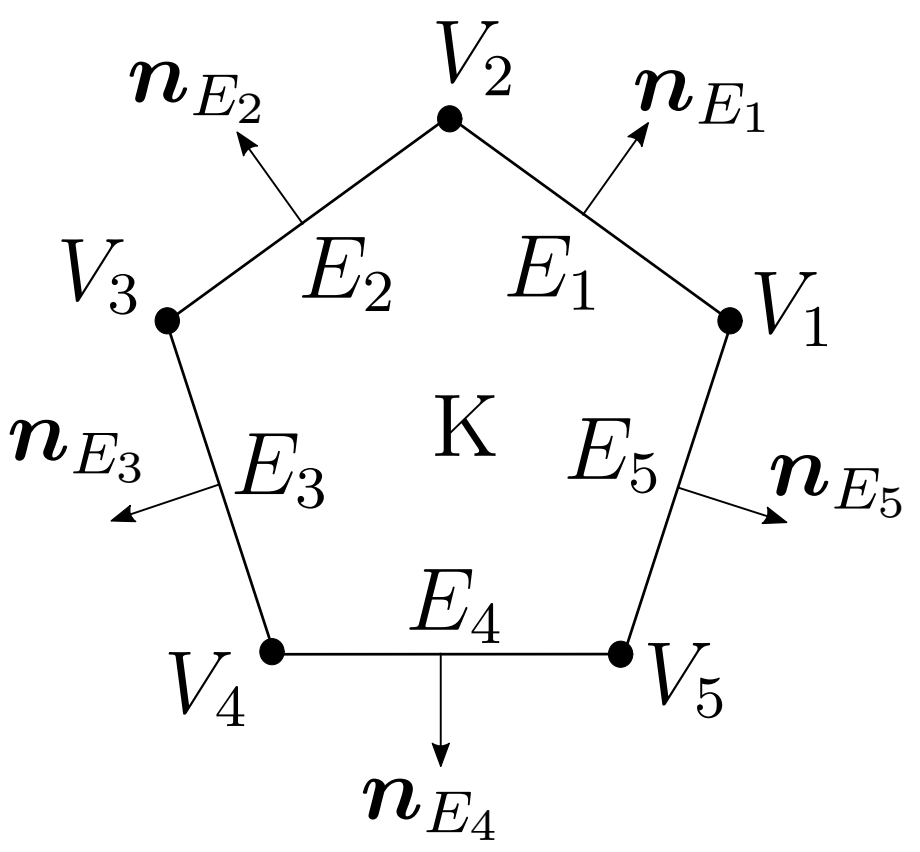

Figure 2: An example of a 5-edge element K.

Furthermore, coefficients of the projection of shape functions are determined due to orthogonal property (13)

$$
\begin{aligned}
& a^K\left(\boldsymbol{m}^\alpha, \Pi \boldsymbol{v}\right) = a^K\left(\boldsymbol{m}^\alpha, \boldsymbol{v}\right) \quad \alpha = 1, 2, ..., 6 \\
\Rightarrow \quad & \sum_{i=1}^{n_V} a^K\left(\boldsymbol{m}^\alpha, \Pi \boldsymbol{\varphi}^i\right) \hat{\boldsymbol{v}}^i = \sum_{i=1}^{n_V} a^K\left(\boldsymbol{m}^\alpha, \boldsymbol{\varphi}^i\right) \hat{\boldsymbol{v}}^i \\
\Rightarrow \quad & \sum_{i=1}^{n_V} \sum_{\beta=1}^{n_k} a^K(\boldsymbol{m}^\alpha, \boldsymbol{m}^\beta)\, \boldsymbol{\pi}^{\beta i} = \sum_{i=1}^{n_V} a^K(\boldsymbol{m}^\alpha, \boldsymbol{\varphi}^i) \\
\Rightarrow \quad & \mathbf{G} \cdot \boldsymbol{\Pi}_* = \mathbf{B}, \qquad (24)
\end{aligned}
$$

where $a^K(\boldsymbol{m}^\alpha, \boldsymbol{m}^\beta)$ and $a^K(\boldsymbol{m}^\alpha, \boldsymbol{\varphi}^i)$ are precisely written in integral form by referring to equation (23)

$$a^K(\boldsymbol{m}^\alpha, \boldsymbol{m}^\beta) = \int_{\Gamma_K} \boldsymbol{m}^\beta \cdot \boldsymbol{\sigma}(\boldsymbol{m}^\alpha) \cdot \boldsymbol{n}\, d\Gamma, \tag{25}$$

$$a^K(\boldsymbol{m}^\alpha, \boldsymbol{\varphi}^i) = \int_{\Gamma_K} \boldsymbol{\varphi}^i \cdot \boldsymbol{\sigma}(\boldsymbol{m}^\alpha) \cdot \boldsymbol{n}\, d\Gamma. \tag{26}$$

As can be seen in equation (24), the bilinear form will be trivial or the system will be undetermined if $\alpha$ equals to 1, 2, or 3. For the case $\alpha = 1, 2$, it means that $\boldsymbol{m}^\alpha$ are the constant elements. For the case $\alpha = 3$, the ansatz is presenting for the rotation part among rigid body translation, shear, axial, and volumetric strains [11]. Consequently, the energy of the bilinear form is zero in both sides. Therefore, we need two additional equations to eliminate this indeterminacy. Those equations are related to the equivalence about average values of an arbitrary function $\boldsymbol{v}$ and its projection $\Pi\boldsymbol{v}$. The first one can be found in (15) and the second one is given by the following form

$$\frac{1}{|K|}\int_{\Omega_K} \nabla \boldsymbol{v}\, d\Omega = \frac{1}{|K|}\sum_{E\in\partial K}\int_E \boldsymbol{v}\otimes \boldsymbol{n}_E\, d\Gamma. \tag{27}$$

Moreover, we need the tensor presentation of the operator $\Pi$. To derive its mathematical expression, we think of a map $V_{h,K} \to V_{h,K}$. Let us consider the Lagrange interpolation like in FEM [41]. We have

$$\Pi\boldsymbol{\varphi}^i = \sum_{I=1}^{n_V} \boldsymbol{\varphi}^I (\hat{\Pi\boldsymbol{\varphi}})^{Ii} = \boldsymbol{\varphi}\cdot \hat{\Pi\boldsymbol{\varphi}}^i. \tag{28}$$

As we known in 21,

$$\Pi\boldsymbol{\varphi}^i = \sum_{\alpha=1}^{n_k} \boldsymbol{m}^\alpha \boldsymbol{\pi}^{\alpha i} = \sum_{\alpha=1}^{n_k}\sum_{I=1}^{n_V} \boldsymbol{\varphi}^I \hat{\boldsymbol{m}}^{I\alpha}\boldsymbol{\pi}^{\alpha i}. \tag{29}$$

From equations (28), (29) and (24), we obtain

$$\begin{aligned} &\sum_{I=1}^{n_V} (\hat{\Pi\boldsymbol{\varphi}})^{Ii} = \sum_{\alpha=1}^{n_k}\sum_{I=1}^{n_V} \hat{\boldsymbol{m}}^{I\alpha}\boldsymbol{\pi}^{\alpha i} \\ \Rightarrow\quad &\mathbf{\Pi} = \mathbf{D}\cdot\mathbf{\Pi}_* = \mathbf{D}\cdot\mathbf{G}^{-1}\cdot\mathbf{B}, \end{aligned} \tag{30}$$

where $\mathbf{D}$ in our case is calculated as

$$\mathbf{D} = \begin{bmatrix} \boldsymbol{m}_1(\mathbf{X}_1) & \boldsymbol{m}_2(\mathbf{X}_1) & \cdots & \boldsymbol{m}_6(\mathbf{X}_1) \\ \boldsymbol{m}_1(\mathbf{X}_2) & \boldsymbol{m}_2(\mathbf{X}_2) & \cdots & \boldsymbol{m}_6(\mathbf{X}_2) \\ \vdots & \vdots & \ddots & \vdots \\ \boldsymbol{m}_1(\mathbf{X}_{n_V}) & \boldsymbol{m}_2(\mathbf{X}_{n_V}) & \cdots & \boldsymbol{m}_6(\mathbf{X}_{n_V}) \end{bmatrix}. \tag{31}$$

Moreover, there is a convenient way to calculate the matrix $\mathbf{G}$ based on the relation between $\mathbf{B}$ and $\mathbf{D}$, the proof of which can be found in [**?** ]

$$\mathbf{G} = \mathbf{B} \cdot \mathbf{D}. \tag{32}$$

### *2.3 Stiffness matrix and force terms*

For two functions $\boldsymbol{u}$, $\boldsymbol{v} \in V_{h,K}$ with $\boldsymbol{v}$ is the arbitrary test function, the local stiffness bilinear form is

$$\begin{aligned} a^K(\boldsymbol{u}, \boldsymbol{v}) &= a^K(\Pi\boldsymbol{u} + (\boldsymbol{u} - \Pi\boldsymbol{u}), \Pi\boldsymbol{v} + (\boldsymbol{v} - \Pi\boldsymbol{v})) \\ &= a^K(\Pi\boldsymbol{u}, \Pi\boldsymbol{v}) + a^K(\boldsymbol{u} - \Pi\boldsymbol{u}, \Pi\boldsymbol{v}) + a^K(\Pi\boldsymbol{u}, \boldsymbol{v} - \Pi\boldsymbol{v}) \\ &\quad + a^K(\boldsymbol{u} - \Pi\boldsymbol{u}, \boldsymbol{v} - \Pi\boldsymbol{v}). \end{aligned} \tag{33}$$

The second and the third term of equation 33 are vanished because of orthogonal property of projection (13). Hence, it yields

$$a^K(\boldsymbol{u}, \boldsymbol{v}) = a^K(\Pi\boldsymbol{u}, \Pi\boldsymbol{v}) + a^K(\boldsymbol{u} - \Pi\boldsymbol{u}, \boldsymbol{v} - \Pi\boldsymbol{v}). \tag{34}$$

We recall that we want our bilinear form $a^K$ to fulfill consistency and stability conditions [1]. To this end, we mimic the identity (34) and choose $a_h^K$ to be in the form

$$a_h^K(\boldsymbol{u}, \boldsymbol{v}) = a^K(\Pi\boldsymbol{u}, \Pi\boldsymbol{v}) + s^K(\boldsymbol{u} - \Pi\boldsymbol{u}, \boldsymbol{v} - \Pi\boldsymbol{v}), \tag{35}$$

where the first term is called the consistency term, and the second term is referred to as the stability term. It is noteworthy to mention that two terminologies stability term and stabilization term in our paper carry different meanings. The bilinear form (35) satisfies the consistency property and stability property as the proof can be found in [1, 3]. The consistency term is explicitly expressed by means of definition of projection $\Pi$ in equations (16),

(20), and (24)

$$
\begin{aligned}
a^K(\Pi \boldsymbol{u}, \Pi \boldsymbol{v}) &= \hat{\boldsymbol{v}} \cdot \mathbf{\Pi}_*^T \cdot a^K(\mathbf{M}^T, \mathbf{M}) \cdot \mathbf{\Pi}_* \cdot \hat{\boldsymbol{u}} \\
&= \hat{\boldsymbol{v}} \cdot \mathbf{B}^T \cdot \mathbf{G}^{-T} \cdot \mathbf{A} \cdot \mathbf{G}^{-1} \cdot \mathbf{B} \cdot \hat{\boldsymbol{u}},
\end{aligned} \tag{36}
$$

where $\mathbf{A}$ is the matrix that is identical to $\mathbf{G}$ with a modification where the first three rows with respect to $\alpha = 1, 2, 3$ are set to zero (rigid body motion and rotation [4, 11]). In addition, with equations (16), (30), the stability term can be fully expressed as follows

$$
\begin{aligned}
a^K(\boldsymbol{u} - \Pi \boldsymbol{u}, \boldsymbol{v} - \Pi \boldsymbol{v}) &= \hat{\boldsymbol{v}} \cdot (\mathbf{I} - \mathbf{\Pi})^T \cdot a^K(\boldsymbol{\varphi}^T, \boldsymbol{\varphi}) \cdot (\mathbf{I} - \mathbf{\Pi}) \cdot \hat{\boldsymbol{u}} \\
&\approx \hat{\boldsymbol{v}} \cdot (\mathbf{I} - \mathbf{D} \cdot \mathbf{G}^{-1} \cdot \mathbf{B})^T \cdot \boldsymbol{S}_*^K \cdot (\mathbf{I} - \mathbf{D} \cdot \mathbf{G}^{-1} \cdot \mathbf{B}) \cdot \hat{\boldsymbol{u}} \\
&= s^K(\boldsymbol{u} - \Pi \boldsymbol{u}, \boldsymbol{v} - \Pi \boldsymbol{v}),
\end{aligned} \tag{37}
$$

where $\mathbf{I}$ in (37) is the identity matrix and $\boldsymbol{S}_*^K$ is the local matrix which ensures the stability. Moreover, the stability part $s^K$ is a symmetric bilinear form that can be chosen so that our bilinear form (35) fulfills the consistency and stability conditions. Hence, $\boldsymbol{S}_*^K$ has a form

$$
\boldsymbol{S}_*^K = \gamma \, \alpha_*^K \mathbf{I}, \tag{38}
$$

where $\gamma > 0$ denotes the multiplicative factor or stabilization parameter [18] to control the magnitude of the material parameter and that can be selected in a certain range depending on a specific problem. The investigation on how to choose it is addressed in [3, 4, 6].

Several stabilization analysis have been conducted such as the identity matrix choice [1], Hourglass stabilization [42], or the correction strain energy [4]. Nevertheless, they are mostly unstable with respect to each ratio scale. Recently, two new stabilization terms used in contact problems where the stabilization is simplified by dropping the contribution of interior degrees of freedom [11], or used in geomechanics [10] are proposed, and they have mathematical proofs. Underlying solutions are stable with respect to isotropic scaling and aspect ratio. The proof of stabilization can be found in [18, 10]. Here, we use a stabilization term adopted from [10], and it is given as follows

$$
\alpha_*^K = \frac{1}{10} \, |K| \, \mathrm{tr}(\boldsymbol{E}) \, \mathrm{tr}((\mathbf{D}_c^T \, \mathbf{D}_c)^{-1}), \tag{39}
$$

where $\boldsymbol{E}$ is the material property, and $\mathbf{D}_c$ are the components of matrix $\mathbf{D}$ in (31) without the constant parts. Finally, we obtain the local stiffness matrix of each polygon $K$ by gathering equations (35), (36), (37), (38), and (39)

$$\mathbf{K}^K = (\underbrace{\mathbf{B}^T \cdot \mathbf{G}^{-T} \cdot \mathbf{A} \cdot \mathbf{G}^{-1} \cdot \mathbf{B}}_{\mathbf{K}_c^K}) + (\underbrace{\mathbf{I} - \mathbf{D} \cdot \mathbf{G}^{-1} \cdot \mathbf{B})^T \cdot \boldsymbol{S}_*^K \cdot (\mathbf{I} - \mathbf{D} \cdot \mathbf{G}^{-1} \cdot \mathbf{B}}_{\mathbf{K}_s^K}), \tag{40}$$

where $\mathbf{K}_c^K$, $\mathbf{K}_s^K$ present for the consistency stiffness matrix and the stability stiffness matrix of each element, respectively. Eventually, the global stiffness matrix is generated by assembling all local stiffness matrices based on their connectivities in analogy to FEM.

In addition, the load term consists of boundary and volume load. While the volume load is assumed to be zero in our case, the boundary load can be calculated as in FEM

$$\begin{aligned} b^K(\boldsymbol{v}) &= \int_{\Gamma_K} \boldsymbol{v} \cdot \boldsymbol{t} \, d\Gamma \\ &= \hat{\boldsymbol{v}} \cdot \int_{\Gamma_K} \boldsymbol{\varphi}^T \cdot \boldsymbol{t} \, d\Gamma. \end{aligned} \tag{41}$$

If we consider the body weight, the load is computed as follows

$$\begin{aligned} \int_{\Omega_K} \boldsymbol{v} \cdot \boldsymbol{b} \, d\Omega &\approx \int_{\Omega_K} \Pi \boldsymbol{v} \cdot \boldsymbol{b} \, d\Omega \\ &= \hat{\boldsymbol{v}} \cdot \int_{\Omega_K} \Pi \boldsymbol{\varphi}^T \cdot \boldsymbol{b} \, d\Omega \\ &= \hat{\boldsymbol{v}} \cdot \mathbf{\Pi}_*^T \cdot \int_{\Omega_K} \mathbf{M}^T \cdot \boldsymbol{b} \, d\Omega. \end{aligned} \tag{42}$$

## 3 Virtual Element Method for Crack Propagation in LEFM

### *3.1 Computation of SIFs*

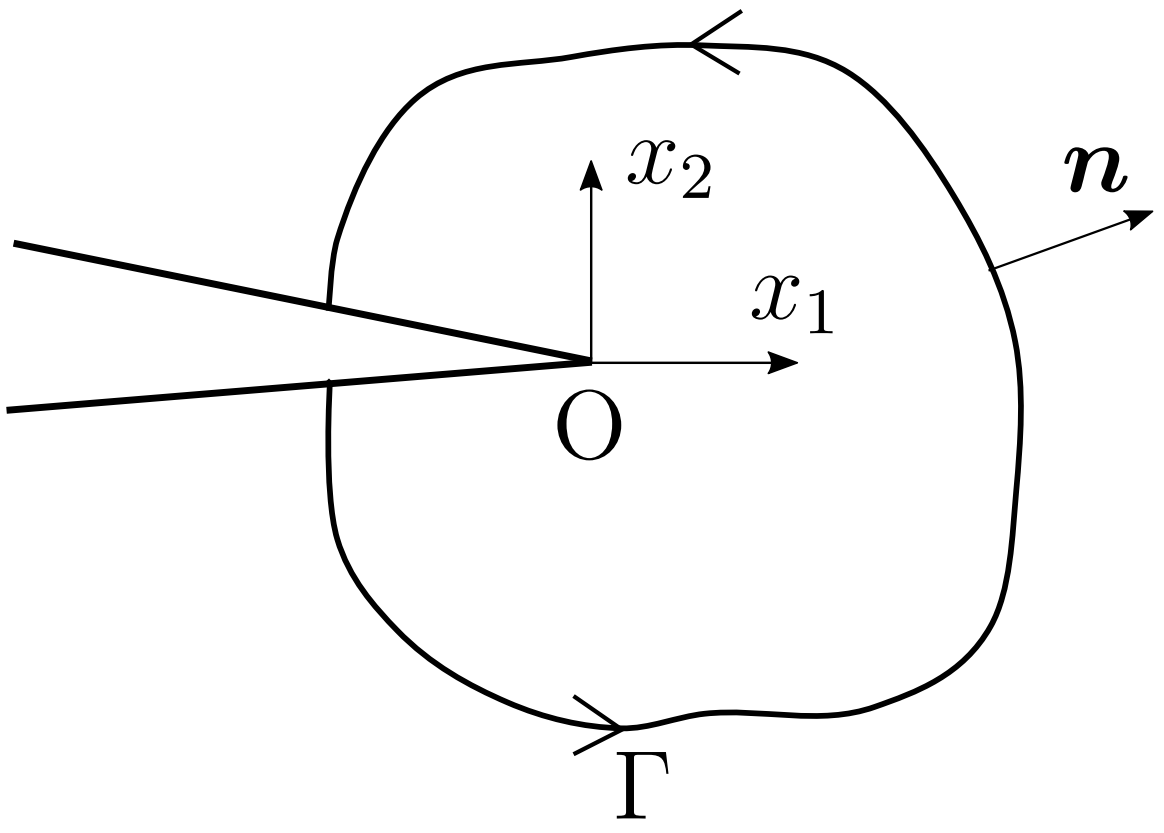

Figure 3: Crack tip coordinates and path of $J$-Integral.

Consider an arbitrary counterclockwise path around a crack tip (Figure 3), Rice et al [39] defined the $J$-integral as follows

$$J = \int_{\Gamma} \left( W dx_2 - T_i \frac{\partial u_i}{\partial x_1} \, ds \right), \tag{43}$$

where $T_i$ is the component of traction vector, $u_i$ is the component of displacement vector, $ds$ is the length increment along the contour $\Gamma$, and $W$ is the strain energy density which is defined as follows

$$W = \int_0^{\varepsilon_{ij}} \sigma_{ij} \, d\varepsilon_{ij}. \tag{44}$$

This $J$-integral is a path independent under small deformation, elastic material behavior and quasi-static assumption without body forces and traction free on the crack surfaces. Nevertheless, this $J$-integral is not suitable for numerical analysis, since it is not feasible to evaluate stress and strain along a vanishingly small contour. The technique developed by Li *et al.* [43] is more applicable in numerical methods as it allows $J$-integral to be calculated over an equivalent area instead of a contour integral. The $J$-integral is now

defined as follows

$$J = \int_{\Omega} (\sigma_{ij} \frac{\partial u_j}{\partial x_1} - W\delta_{1i}) \frac{\partial q}{\partial x_i} \, d\Omega, \tag{45}$$

where $q$ can be seen as a sufficiently smooth weighting function in area $\Omega$ that is unity on an open set containing the crack tip and vanishes on an outer contour as shown in Figure 4.

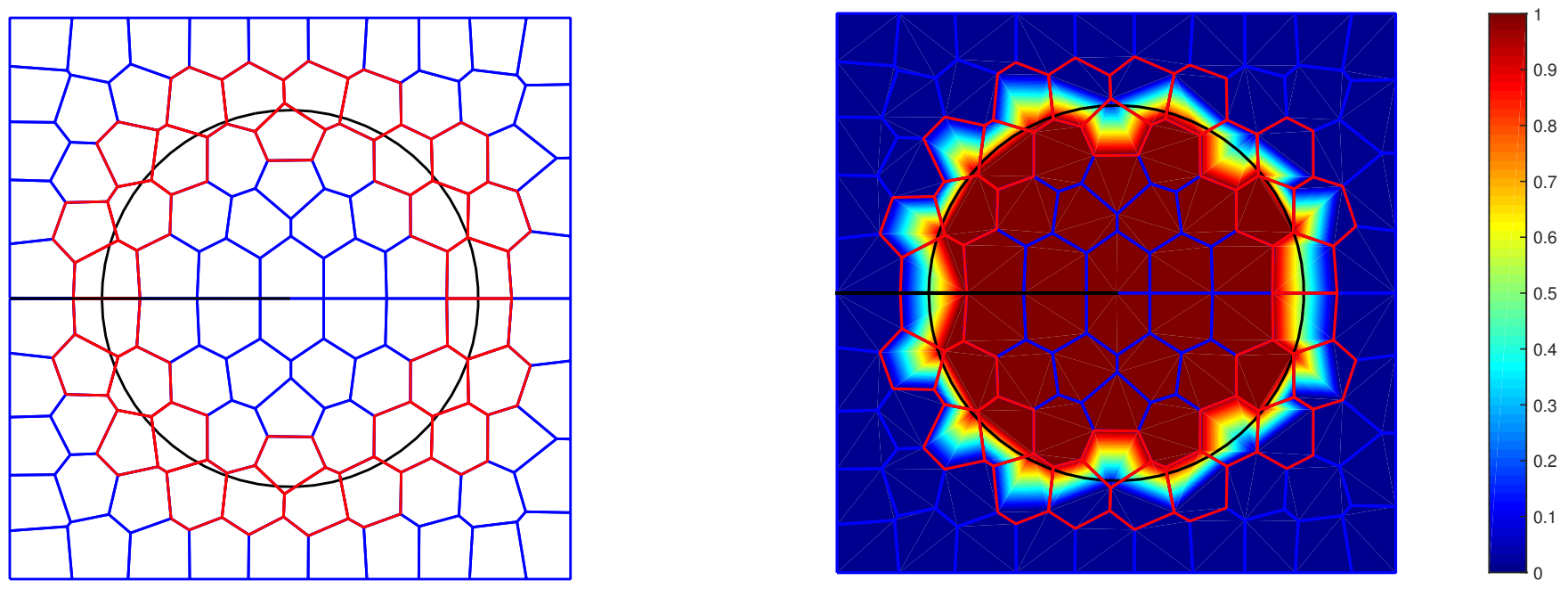

Figure 4: $J$ domain (red elements) and weighting function $q$ in a typical polygonal mesh.

Based on the $J$-integral, an interaction integral in mixed-mode [40] is used to calculate the stress intensity factors. Two states, a present state and an auxiliary state, which are denoted by superscripts (1) and (2), respectively, are used in the integration. Herein, the auxiliary state is chosen as the asymptotic fields of a crack body for modes $I$ or $II$. The final form for the interaction integral [44], [45] for regular elements is written as follows

$$I^{(1,2)} = \int_{\Omega} F_j(x_1, x_2) \frac{\partial q}{\partial x_j} \, d\Omega, \tag{46}$$

where $F_j(x_1, x_2)$ is given in the following

$$F_j(x_1, x_2) = \sigma_{ij}^{(1)} \frac{\partial (u_i^{(2)})}{\partial x_1} + \sigma_{ij}^{(2)} \frac{\partial (u_i^{(1)})}{\partial x_1} - W^{(1,2)} \delta_{1j}, \tag{47}$$

and the interaction strain energy $W^{(1,2)}$ is given in the following

$$W^{(1,2)} = \sigma_{ij}^{(1)} \varepsilon_{ij}^{(2)} = \sigma_{ij}^{(2)} \varepsilon_{ij}^{(1)}. \tag{48}$$

This interaction integral, however, is not suitable for VEM because the integral in VEM has to be evaluated over the polygons. Discretizing the domain corresponding to $J$-integral curve into small elements and applying the Gaussian theorem to transform the area integral to boundary integral, we obtain

$$I^{(1,2)} = \sum_{K=1}^{N_J} \left( \int_{\Gamma_K} F_j(x_1, x_2) q \, n_j \, d\Gamma - \int_{\Omega_K} F_{j,j}(x_1, x_2) q \, d\Omega \right), \tag{49}$$

where $N_J$ is the number of elements involving in J-domain. Due to the absence of body force and $W^{(1,2)} = \frac{1}{2}\sigma_{ij}^{(1)}\varepsilon_{ij}^{(2)} + \frac{1}{2}\sigma_{ij}^{(2)}\varepsilon_{ij}^{(1)}$, the second part becomes zero. Thus, the interaction integral over the boundaries of elements is left. In addition, thanks to the projection operator, because of the presence of that, weighting function $q$, the components of both displacement field $u_i^{(1)}$ and stress field $\sigma_{ij}^{(1)}$ in the present state can be replaced by their projections in VEM. The interaction integral (46) now has the final form, that is

$$I^{(1,2)} = \sum_{K=1}^{N_J} \int_{\Gamma_K} \left[ \Pi\sigma_{ij}^{(1)} \frac{\partial (u_i^{(2)})}{\partial x_1} + \sigma_{ij}^{(2)} \frac{\partial (\Pi u_i^{(1)})}{\partial x_1} - W^{(1,2)} \delta_{1j} \right] \Pi q \, n_j \, d\Gamma. \tag{50}$$

Finally, SIFs for mode $I$ and mode $II$ are given based on interaction integral

$$K_I^{(1)} = \frac{E^*}{2} I^{(1,\text{Mode } I)}, \tag{51}$$

$$K_{II}^{(1)} = \frac{E^*}{2} I^{(1,\text{Mode } II)}, \tag{52}$$

where $E^*$ is defined based on Young's modulus $E$ and poisson ratio $\nu$

$$E^* = \begin{cases} E & \text{plane stress} \\ \frac{E}{1-\nu^2} & \text{plane strain.} \end{cases} \tag{53}$$

### *3.2 Adaptivity and crack propagation*

#### *3.2.1 Superconvergent patch recovery (SPR)*

It is well known that the stresses at certain points within an element can be used for stress recovery for superconvergence [35, 36]. It means the convergence rate of superconvergent stress field is equivalent to that of the displacement field in odd order approximation, i.e the convergent order in $O(h^{k+1})$. For example, the order of convergence for linear approximate stress field could reach to $O(h^2)$. While in even order approximation, the superconvergent stress can reach to $O(h^{k+2})$. It has been shown that in the context of FEM, the stress at the Gauss integration points should possess the superconvergence property. In VEM, the Gauss points is not a mandatory choice using the polygonal element. In fact, the stress field can be recovered at certain points known as sampling points, which are closed to Gauss points. Without loss of generality we assume that the stress denoted by $\hat{\boldsymbol{\sigma}}$ is superconvergent at certain sampling points. Therefore, it shall yield the superconvergent stress field $\boldsymbol{\sigma}^*$ for all points within a patch (Figure 5).

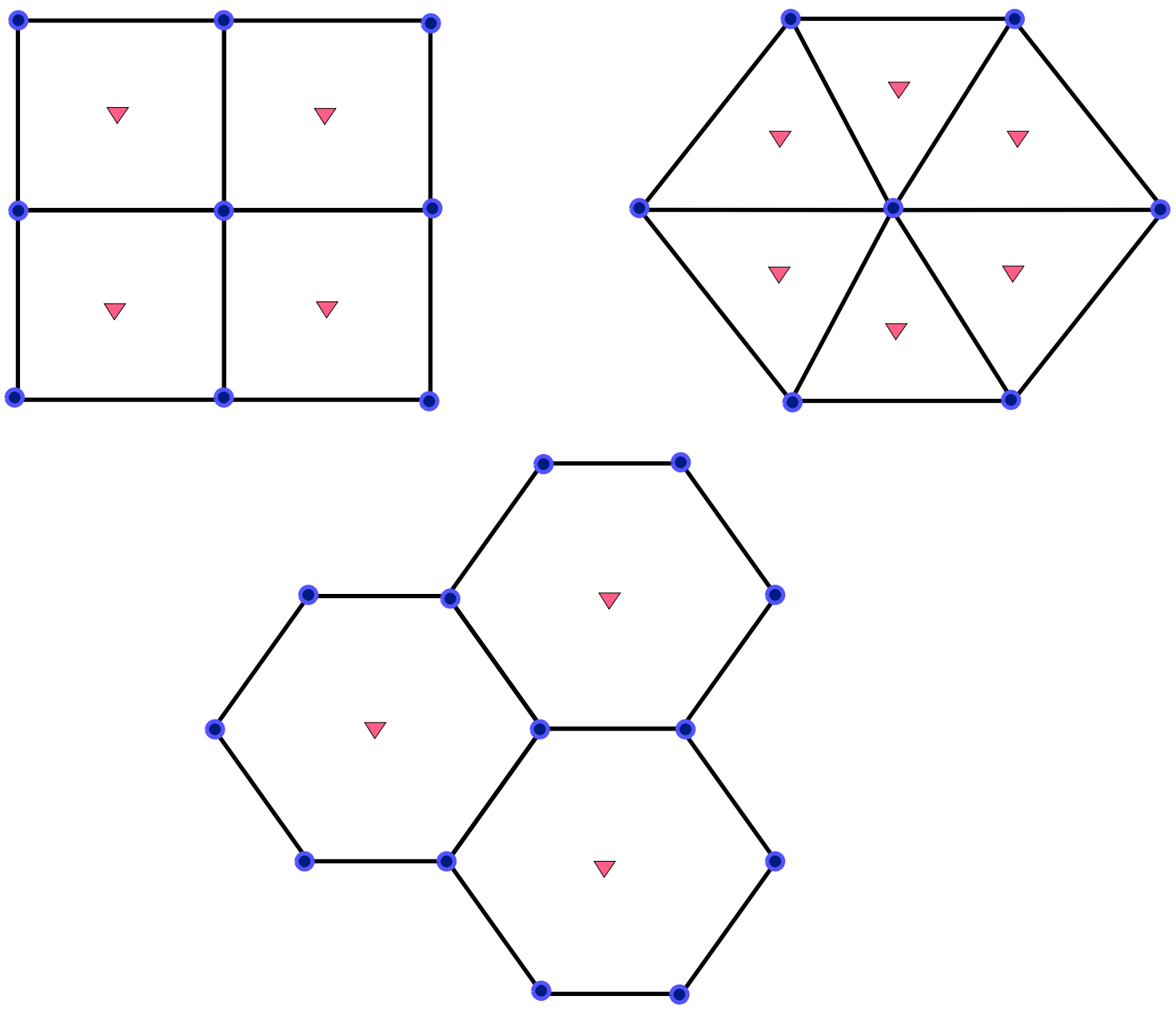

Figure 5: Typical patches for Q4 elements, T3 elements, polygonal elements. The recovered nodes are presented by blue circles, and sampling points (centroids) are denoted by red triangles.

First, the superconvergent stress at the nodal point is approximated with the identical order of Ansatz functions for the displacement, that is

$$\tilde{\sigma}_i^* = \boldsymbol{p}(x, y)\, \boldsymbol{a}_i, \tag{54}$$

in which

$$\boldsymbol{p}(x, y) = [\, 1 \;\, x \;\, y \,], \tag{55}$$

$$\boldsymbol{a}_i = [\, a_1 \;\, a_2 \;\, a_3 \,]. \tag{56}$$

With $n$ distributed sampling points, which are centroids of polygons, we consider the minimization problem of the error $\Pi$ as illustrated in Figure 6 to find the curve fitting

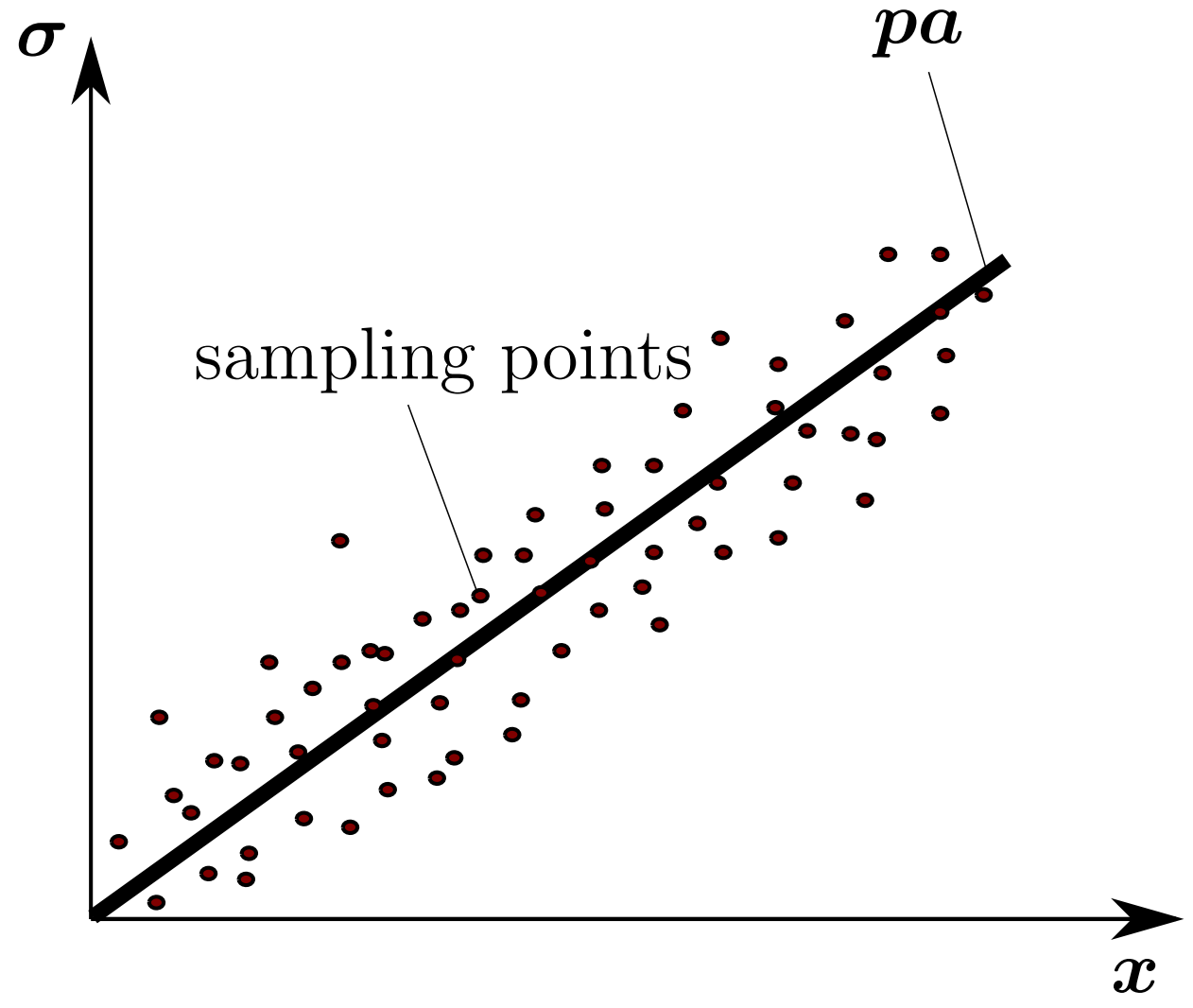

Figure 6: A sketch shows the superconvergent stress line and surrounding sampling points.

$$\Pi = \frac{1}{2} \sum_{k=1}^{n} \left[\hat{\sigma}_i(x_k, y_k) - \boldsymbol{p}_k \boldsymbol{a}_i\right]^2. \tag{57}$$

We minimize that least square functional

$$\frac{\partial \Pi}{\partial \boldsymbol{a}} = \sum_{k=1}^{n} \boldsymbol{p}_k^T \left[\hat{\sigma}_i(x_k, y_k) - \boldsymbol{p}_k \boldsymbol{a}_i\right] = 0. \tag{58}$$

Equation (58) represents a linear regression, so we simply solve linear equa-tion to find the coefficient $\boldsymbol{a}_i$ as follows

$$\boldsymbol{a}_i = \boldsymbol{A}^{-1}\boldsymbol{b}_i, \tag{59}$$

where $\boldsymbol{A} = \sum_{k=1}^{n} \boldsymbol{p}_k^T \boldsymbol{p}_k$, $\boldsymbol{b}_i = \sum_{k=1}^{n} \boldsymbol{p}_k^T \hat{\sigma}_i(x_k, y_k)$, and $\hat{\sigma}_i$ are the components of stress at sampling points. When we have the coefficient $\boldsymbol{a}_i$, the recovered solution of every node can be obtained based on the equation (54). In order to develop the recovery solution for entire element domain in VEM, thanks to the projection operator, we project the Lagrange shape functions onto the polynomial space, and it is used to approximate the superconvergent stress for polygonal elements

$$\boldsymbol{\sigma}^* = \sum_{i=1}^{n_V} \Pi \boldsymbol{\varphi}^i (\tilde{\boldsymbol{\sigma}}^*(x_V, y_V))^i. \tag{60}$$

### *3.2.2 A posteriori error estimation*

Let us briefly go through some criteria convergence and the priori error estimates in our measurement. Let $\boldsymbol{u}^h$ be the VEM solution and $\boldsymbol{u}$ be the analytical solution for the displacement. The total error of the primary field, so called $L^2$ error norm, is measured in relative form as follows

$$e_{L^2} = \frac{||\boldsymbol{u} - \boldsymbol{u}^h||}{||\boldsymbol{u}||} = \frac{\left(\int_\Omega \left[\boldsymbol{u} - \boldsymbol{u}^h\right]^2 d\Omega\right)^{\frac{1}{2}}}{\left(\int_\Omega \boldsymbol{u}^2 \, d\Omega\right)^{\frac{1}{2}}}. \tag{61}$$

Regarding the $H^1$ seminorm or energy error norm, the VEM solution is denoted by $\boldsymbol{\sigma}^h$ which is an approximation of the exact solution $\boldsymbol{\sigma}$ for the stress. The relative error in terms of energy norm is defined as follows

$$e_{H^1} = \frac{\left(\frac{1}{2}\int_\Omega \left[\boldsymbol{\sigma} - \boldsymbol{\sigma}^h\right]^T \boldsymbol{D}^{-1} \left[\boldsymbol{\sigma} - \boldsymbol{\sigma}^h\right] \, d\Omega\right)^{\frac{1}{2}}}{\left(\frac{1}{2}\int_\Omega \boldsymbol{\sigma}^T \boldsymbol{D}^{-1} \boldsymbol{\sigma} \, d\Omega\right)^{\frac{1}{2}}}, \tag{62}$$

where $\boldsymbol{D}$ is the constitutive matrix. For most of analysis, because the analytical solution is not always available, we adopt the recovery solution obtained by (60) as a substitution for the exact one. Thus, the error in this case, so called posterior error, can be calculated in the following

$$||e^*||_{en} = \left( \frac{1}{2} \int_{\Omega} \left[ \boldsymbol{\sigma}^* - \boldsymbol{\sigma}^h \right]^T \boldsymbol{D}^{-1} \left[ \boldsymbol{\sigma}^* - \boldsymbol{\sigma}^h \right] \, d\Omega \right)^{\frac{1}{2}}, \tag{63}$$

and the relative error in terms of the energy error norm of the entire domain is obtained by discretized error of elements

$$\eta = \frac{\left( \frac{1}{2} \sum \int_{\Omega_K} \left[ \boldsymbol{\sigma}^* - \boldsymbol{\sigma}^h \right]^T \boldsymbol{D}^{-1} \left[ \boldsymbol{\sigma}^* - \boldsymbol{\sigma}^h \right] \, d\Omega \right)^{\frac{1}{2}}}{\left( \frac{1}{2} \sum \int_{\Omega_K} (\boldsymbol{\sigma}^*)^T \boldsymbol{D}^{-1} (\boldsymbol{\sigma}^*) \, d\Omega \right)^{\frac{1}{2}}}. \tag{64}$$

The $\eta$ herein becomes an error estimator to control the desired mesh, and the global indicator $||e^*||_{en}$ probably provide us the local indicator to determine elements which will be marked based on Dörfler criterion [46] in adaptive schemes.

#### *3.2.3 Adaptive mesh and crack propagation*

Using uniform fine mesh is computationally expensive to model crack propagation, and in this case adaptive mesh refinement is more efficient in terms of the computational cost and solution accuracy. The mesh size should be adjusted locally and refined automatically and adaptively. Therefore, one needs an effective error estimator or indicator to guide the mesh refinement process and the error indicator defined in Section 3.2.2 will be used. The algorithms used to split marked elements in our work are newest vertex bisection (Figure 7a), polyTree (Figure 7c) and a typical VEM refinement named midPoint refinement (Figure 7b). It is worth mentioning that in newest vertex bisection, triangular elements are refined in such a way to ensure shape regularity and conformity properties. However, the cost for refinement in this scheme is more expensive than the others because not only the element itself, but also the neighboring elements and hanging nodes which are located in the middle of edges are refined. Whereas polyTree or midPoint only refines the marked elements and disregards other elements, although the conformity property is lost by the appearance of exposed mid-side nodes in these two schemes.

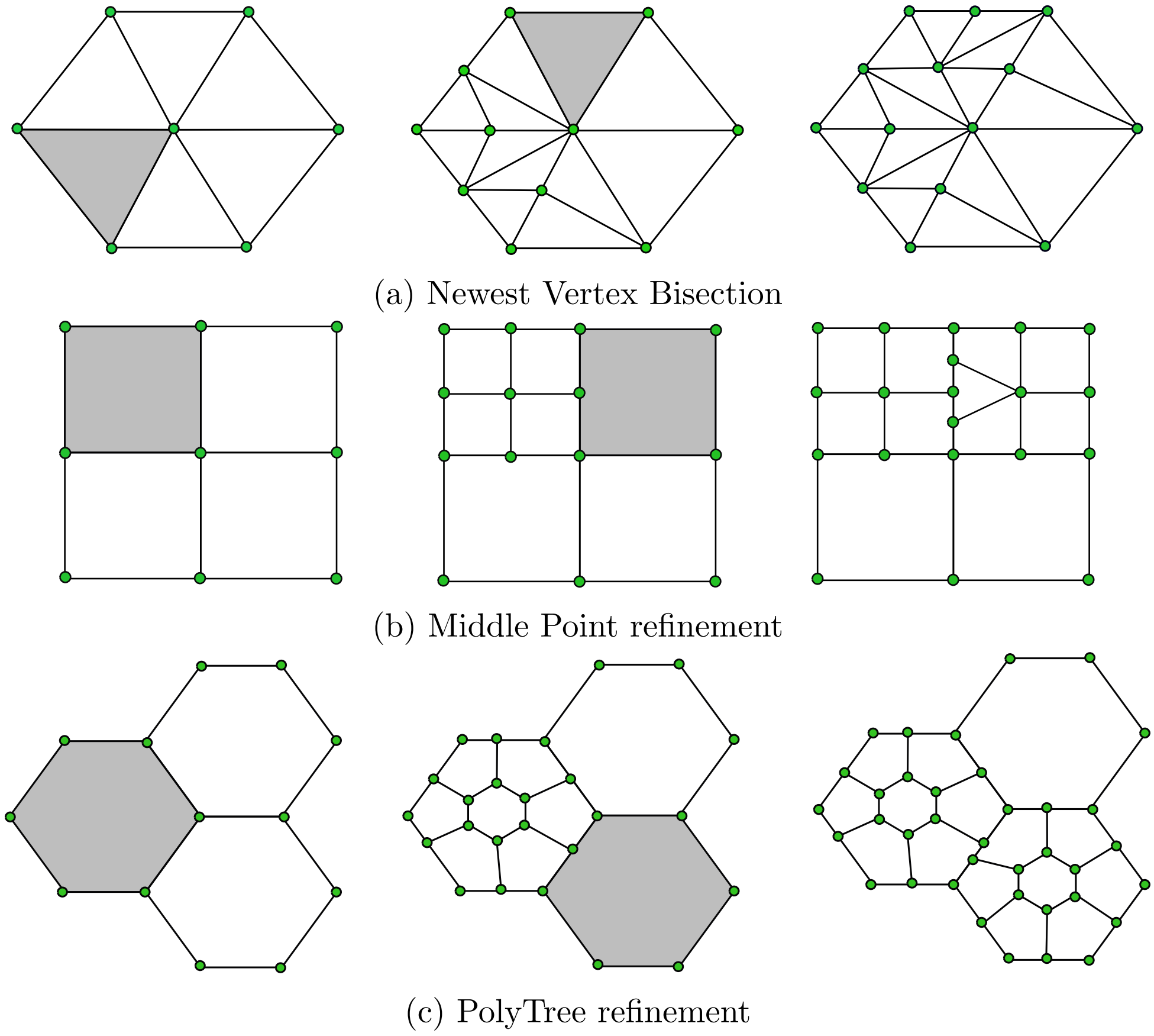

Figure 7: The refinement schemes for three typical VEM meshes in 2 steps. The refined elements are marked in gray color.

Standard finite elements are $C^0$ and shape functions usually have to fulfill four conditions: partition of unity, Kronecker-delta, linear completeness and inter-element compatibility. In PFEM when hanging nodes appear in the refinement, e.g. quadTree or polyTree, the inter-element compatibility condition is violated since $C^0$ continuity is lost along the edges containing mid-side nodes (Figure 8). A modification on formulation of shape functions therefore has to be precisely taken into account. The boundary containing a mid-side node must be parametrized to form an additional shape function corresponding to the hanging node to retain the $C^0$ continuity [47]. Same issue also occurs in XPFEM. In PSBFEM the hanging nodes are not involved because the method automatically capture crack propagation by inducing minimal changes to the global mesh. However, the procedure is highly demanding since much efforts need to be paid to the local mesh refinement and subdomain remerging to track the crack growth path.

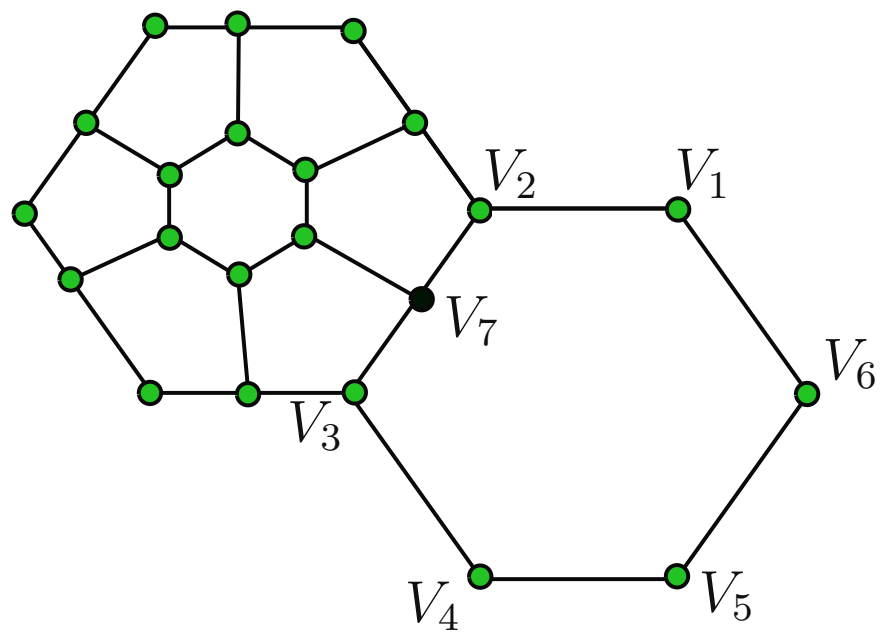

Figure 8: A refined element exposes the hanging node ($V_7$).

The complication of shape functions due to sudden presence of hanging nodes no longer exists in VEM. The steps to bypass this problem are briefly outlined as follows:

- The elements violating the inter-element compatibility condition are firstly removed and then replaced by new polygonal element where the number of nodes has increased due to the additional hanging nodes.

- To assure the efficient approximation, the hanging nodes are moved to center areas and the stabilization terms are chosen so that they are stable with respect to isotropic scaling and aspect ratio if the hanging nodes closely approach the existing nodes, see Figure 9.

For the first step, the computation of stiffness matrix remains simple though there are more nodes. This is because an explicit form of shape functions is not required in VEM. Instead, the coefficients of the projection of shape functions by equation (24) are computed and with these coefficients the consistency part of local stiffness matrix is computable in equation (36). Regarding the stability part, the information which is, in particular, the physical coordinates of both added nodes (hanging nodes) and existing nodes are integrated in matrices $\boldsymbol{B}$, $\boldsymbol{G}$ and $\boldsymbol{D}$ as shown in equation (37).

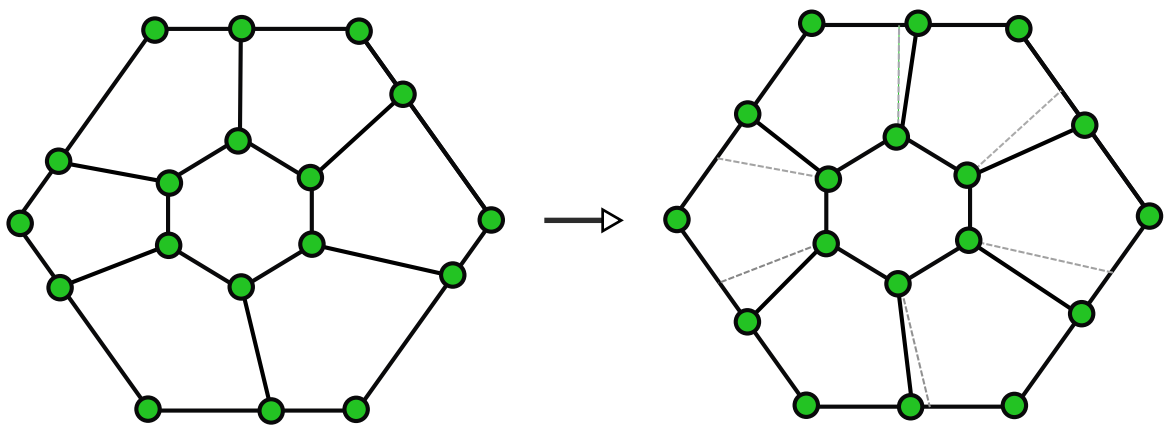

Figure 9: The step to make "nicer" elements.

Regarding the development of crack based on the idea of maximum circumferential stress criterion [48], the propagation angle $\theta$ is defined between the crack line and the crack growth direction, and the positive value is assumed in the counter-clockwise. It is derived as follows

$$\theta = 2\tan^{-1}\left(\frac{K_I}{4K_{II}} - \frac{1}{4}\sqrt{\left(\frac{K_I}{K_{II}}\right)^2 + 8}\right) \quad \text{for} \quad K_{II} > 0, \tag{65}$$

$$\theta = 2\tan^{-1}\left(\frac{K_I}{4K_{II}} + \frac{1}{4}\sqrt{\left(\frac{K_I}{K_{II}}\right)^2 + 8}\right) \quad \text{for} \quad K_{II} < 0. \tag{66}$$

If the equivalent mode $I$ SIF reaches the fracture toughness, the crack will initiate. This equivalent SIF is given by

$$K_{eq} = K_I \cos^3\frac{\theta}{2} - \frac{3}{2}K_{II}\cos\frac{\theta}{2}\sin\theta. \tag{67}$$

Finally, an algorithm of adaptive mesh refinement is briefly summarized in the flowchart below

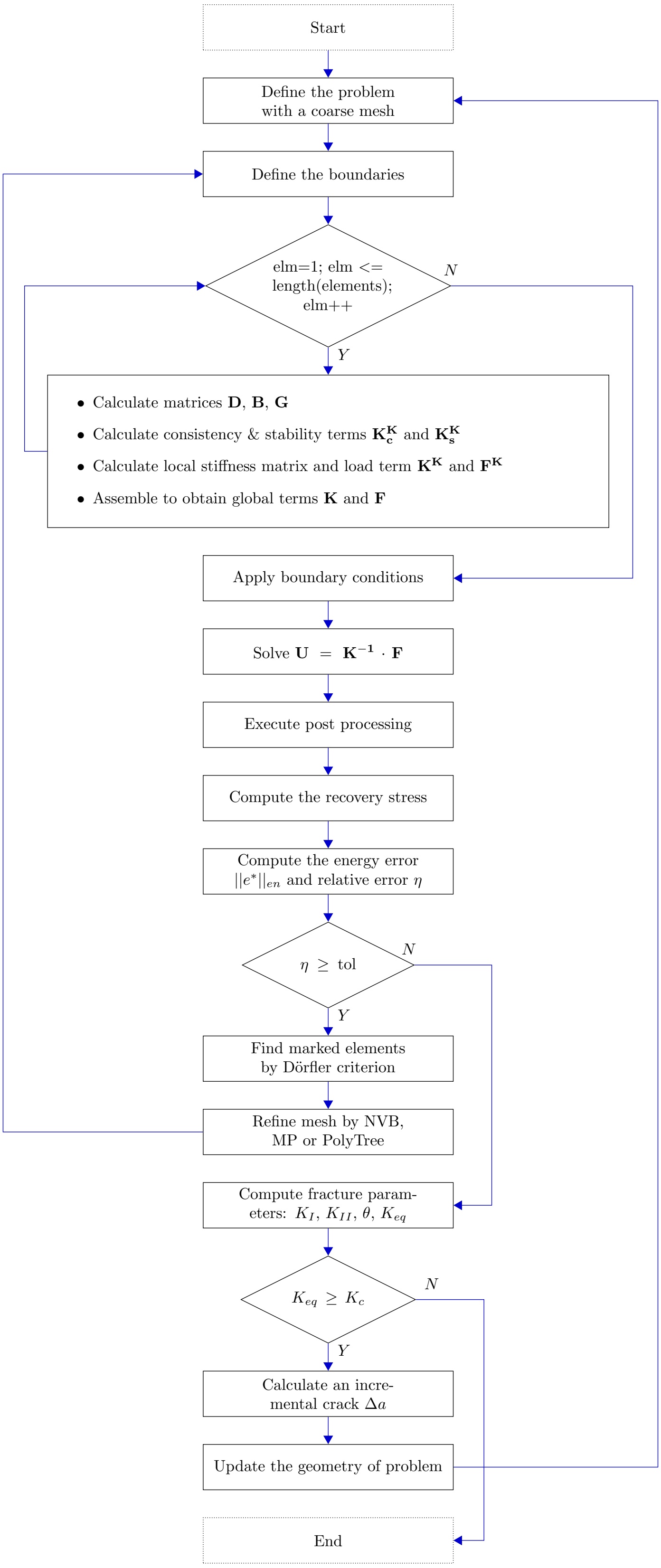

Figure 10: Adaptive mesh refinement for crack propagation algorithm.

## 4 Numerical examples

### *4.1 The Timoshenko beam*

Although we will demonstrate VEM for crack problems, we will first investigate a continuum without cracks and compare the convergence rate with FEM. To this end, let us take the Timoshenko beam as an example [49]. Consider a 2D cantilever beam of length $L$, height $D$, and unit thickness subjected to a parabolic shear load at the free end. The displacement boundary conditions, moreover, are prescribed on the left end of the beam as shown in Figure 11.

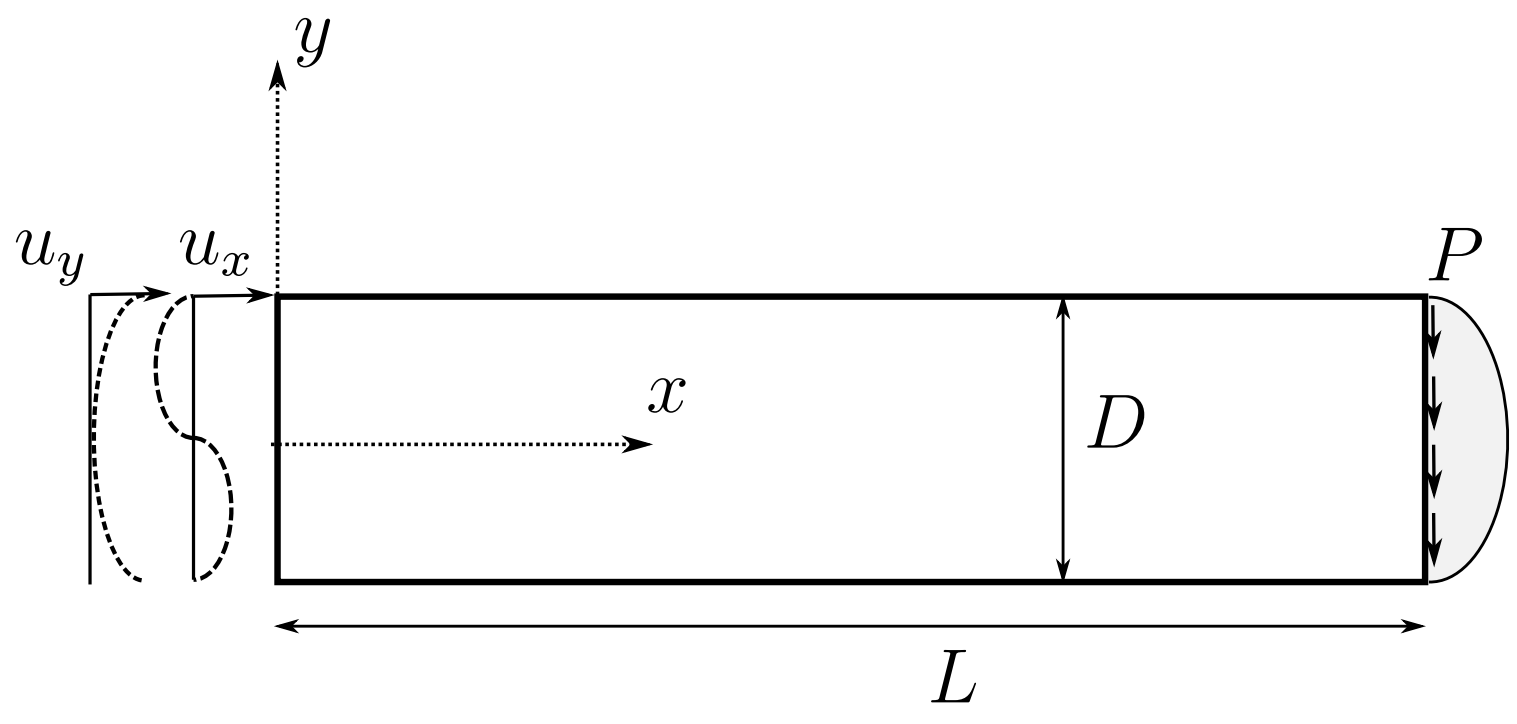

Figure 11: Timoshenko beam: geometry and boundary conditions

The beam is considered in plane stress state. The geometry is formed of length $L = 16$m and height $D = 4$m. The material properties are $E = 1 \times 10^6 \text{N/m}^2$, $\nu = 0.3$, and $P = 1000$N. Timoshenko and Goodier [49] show that the stress field in the cantilever is given by

$$\begin{aligned}
\sigma_{xx} &= \frac{P(L-x)y}{I}, \\
\sigma_{yy} &= 0, \\
\tau_{xy} &= -\frac{P}{2I}\left(\frac{D^2}{4} - y^2\right),
\end{aligned} \tag{68}$$

and the exact solution for displacement field is given by

$$
\begin{aligned}
u_x &= \frac{P\,y}{6EI}\left[(6L-3x)\,x+(2+\nu)\left(y^2-\frac{D^2}{4}\right)\right],\\
u_y &= -\frac{P}{6EI}\left[3\nu y^2\,(L-x)+(4+5\nu)\,\frac{D^2x}{4}+(3L-x)\,x^2\right],
\end{aligned}
\tag{69}
$$

where $I = D^3/12$ is the moment of inertia by assuming unit thickness in 2D.

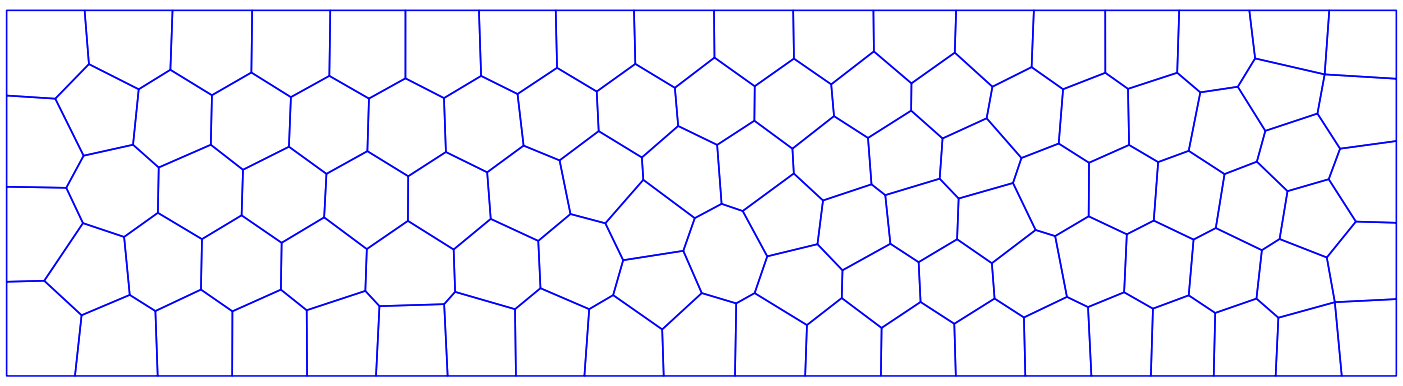

(a) Voronoi mesh (400 D.O.F)

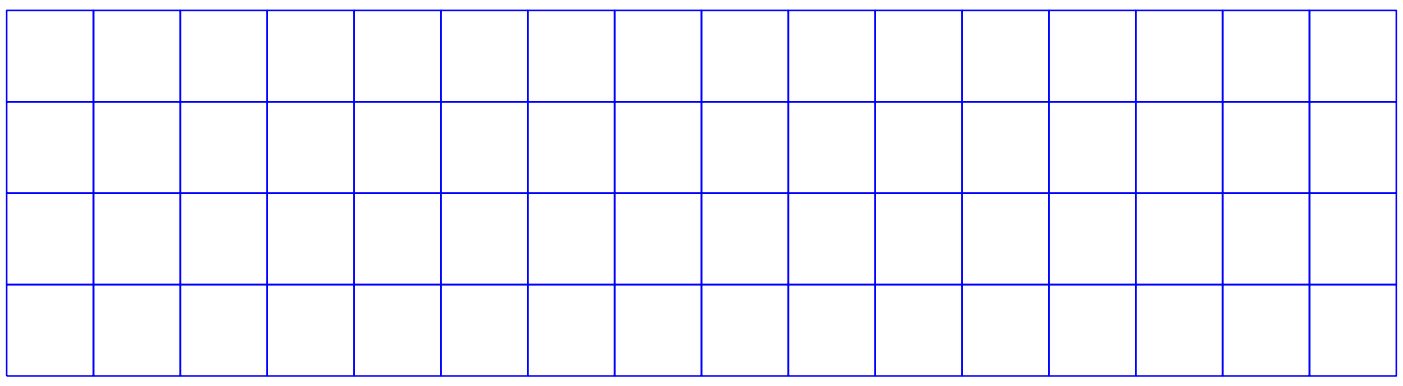

(b) Q4 mesh (170 D.O.F)

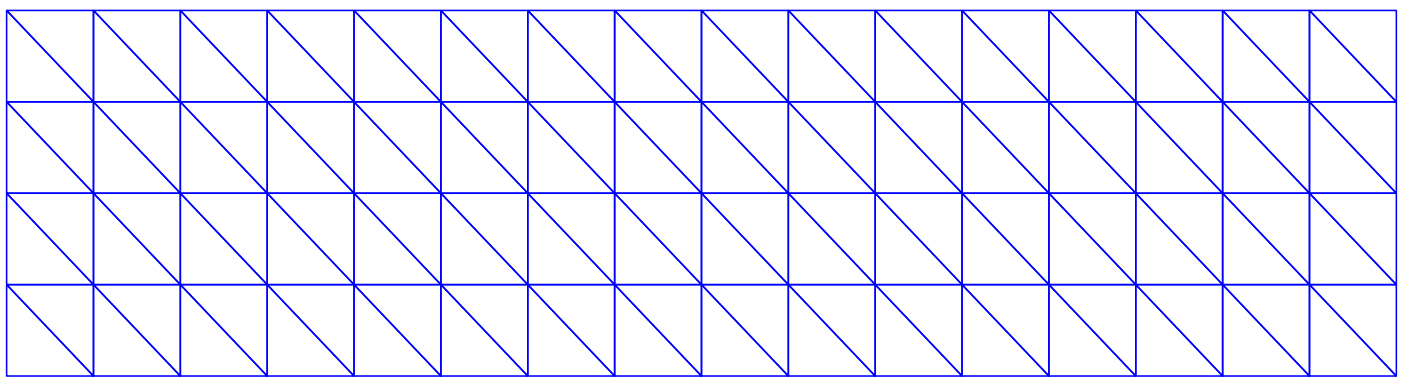

(c) T3 mesh (170 D.O.F)

Figure 12: Three coarse meshes for Timoshenko beam.

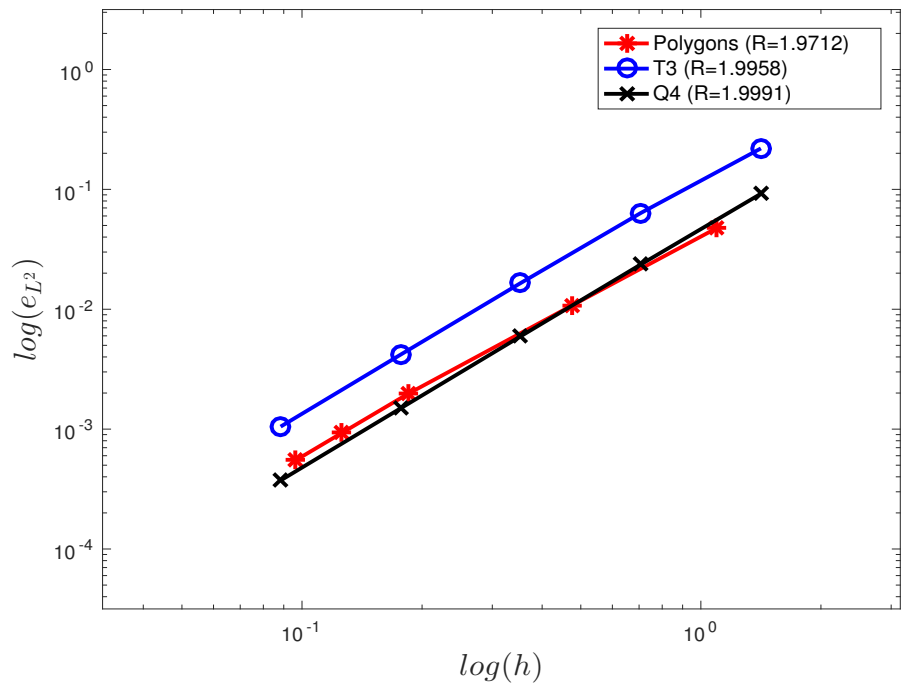

(a)

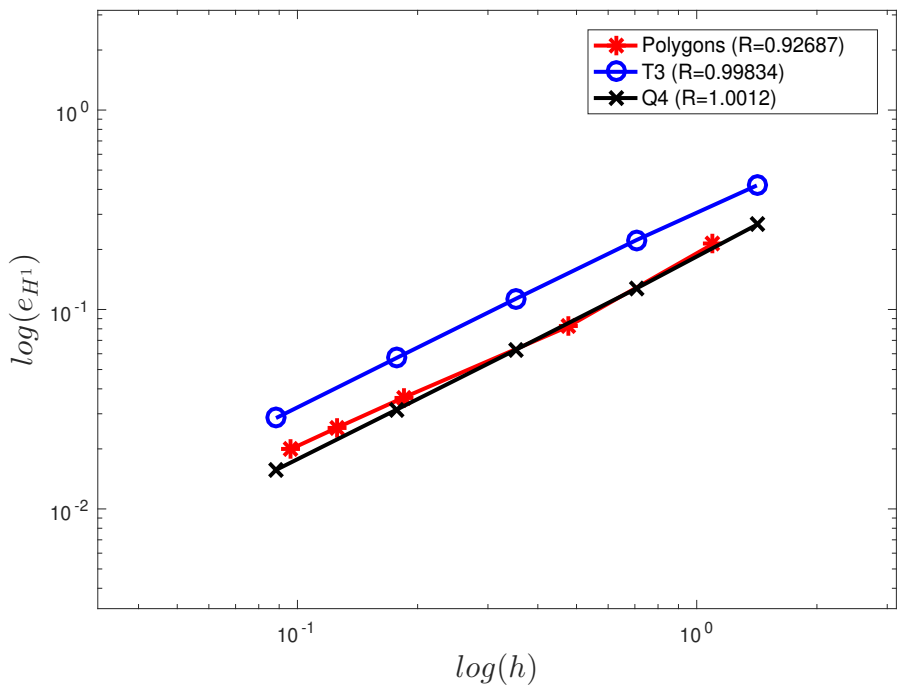

(b)

Figure 13: Convergence rate of $L^2$-norm of error in displacement (a) and energy error norm (b) in Timoshenko beam.

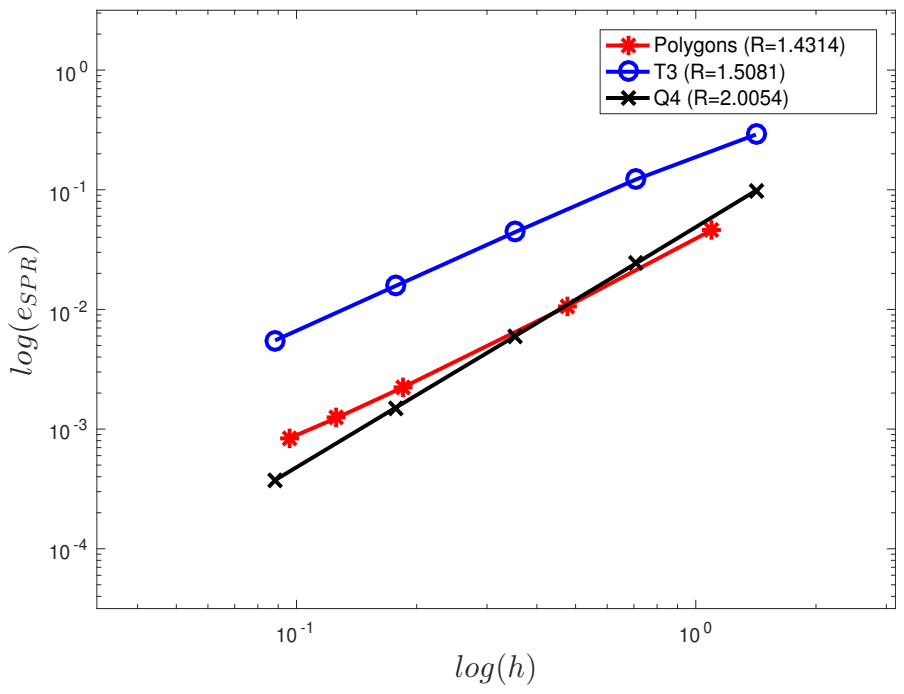

Figure 14: Convergence rate of energy error norm with superconvergent stress in Timoshenko beam.

In this analysis the convergence rate of three types of decompositions Voronoi, Q4, and T3 elements as shown in Figure 12 are investigated. To this end, we discretize the problem domain into five sets of meshes with descending element size $h$ for each type of element. The convergence rates of displacement (61) and energy (62) are progressively measured. In order to calculate the integration in two formulas, we triangulate the polygonal domains and apply the quadrature rules.

As we can see from the graph in Figure 13, the convergence rate of displacement (13a) and energy (13b) in our test almost reach a rate of 2.0 and 1.0, respectively. That means the $L^2$-norm of the error in displacement agrees with convergence rate of order $O(h^2)$ and $H^1$-seminorm of the error in energy agrees with convergence rate of order $O(h)$. The results are significantly consistent with the optimal convergence rate theoretically mentioned in [1, 3]. Moreover, we numerically examine the convergence rate with the superconvergent stress derived in equation (60). Figure 14 shows the rate measured by

$$e_{SPR} = \frac{\left(\frac{1}{2}\int_\Omega \left[\boldsymbol{\sigma} - \boldsymbol{\sigma}^*\right]^T \boldsymbol{D}^{-1} \left[\boldsymbol{\sigma} - \boldsymbol{\sigma}^*\right] \, d\Omega\right)^{\frac{1}{2}}}{\left(\frac{1}{2}\int_\Omega (\boldsymbol{\sigma})^T \boldsymbol{D}^{-1} (\boldsymbol{\sigma}) \, d\Omega\right)^{\frac{1}{2}}}.$$

The convergence rates of meshes discretized in T3, Q4, and polygonal elements are from 1.4 to 2.0. The rate in Q4 obviously shows a true superconvergence while the rates in T3 and Voronoi are still reaching that superconvergence rate because the centroids of elements involved in the formulation are close to the sampling Gauss points for Q4, but for T3 and Voronoi.

### *4.2 Infinite plate with a horizontal crack*

Consider an infinite plate with a center straight crack of length $2a$ under a vertical pulling test with uniform stress field $\sigma$ as described in [50]. Analytical solutions in the vicinity of the crack tip for displacement and stress in polar coordinates with root coordinate at the crack tip are given by:

$$\begin{aligned}
u_x(r,\theta) &= \frac{2(1+\nu)}{\sqrt{2\pi}} \frac{K_I}{E} \sqrt{r} \cos\frac{\theta}{2} \left(2 - 2\nu - \cos^2\frac{\theta}{2}\right), \\
u_y(r,\theta) &= \frac{2(1+\nu)}{\sqrt{2\pi}} \frac{K_I}{E} \sqrt{r} \sin\frac{\theta}{2} \left(2 - 2\nu - \cos^2\frac{\theta}{2}\right), \\
\sigma_{xx}(r,\theta) &= \frac{K_I}{\sqrt{2\pi r}} \cos\frac{\theta}{2} \left(1 - \sin\frac{\theta}{2} \sin\frac{3\theta}{2}\right), \\
\sigma_{yy}(r,\theta) &= \frac{K_I}{\sqrt{2\pi r}} \cos\frac{\theta}{2} \left(1 + \sin\frac{\theta}{2} \sin\frac{3\theta}{2}\right), \\
\sigma_{xy}(r,\theta) &= \frac{K_I}{\sqrt{2\pi r}} \sin\frac{\theta}{2} \cos\frac{\theta}{2} \cos\frac{3\theta}{2},
\end{aligned}$$

where $K_I = \sigma\sqrt{\pi a}$ is the stress intensity factor (SIF), $\nu$ is Poisson's ratio and $E$ is Young's modulus. As the analytical solutions are not available in the entire domain, we investigate solutions near the crack tip by considering the square ABCD of length $10\,\text{mm} \times 10\,\text{mm}$, $a = 100\,\text{mm}$, $\nu = 0.3$, $\sigma =$ $10^4\,\mathrm{N/mm^2}$, $E = 10^7\,\mathrm{N/mm^2}$, and the modeled crack length is 5 mm. Plane strain state is assumed in this problem. Dirichlet boundary conditions are applied on the DC, AB, AD edges and Neumann boundary conditions are applied on the BC edge as shown in Figure 15.

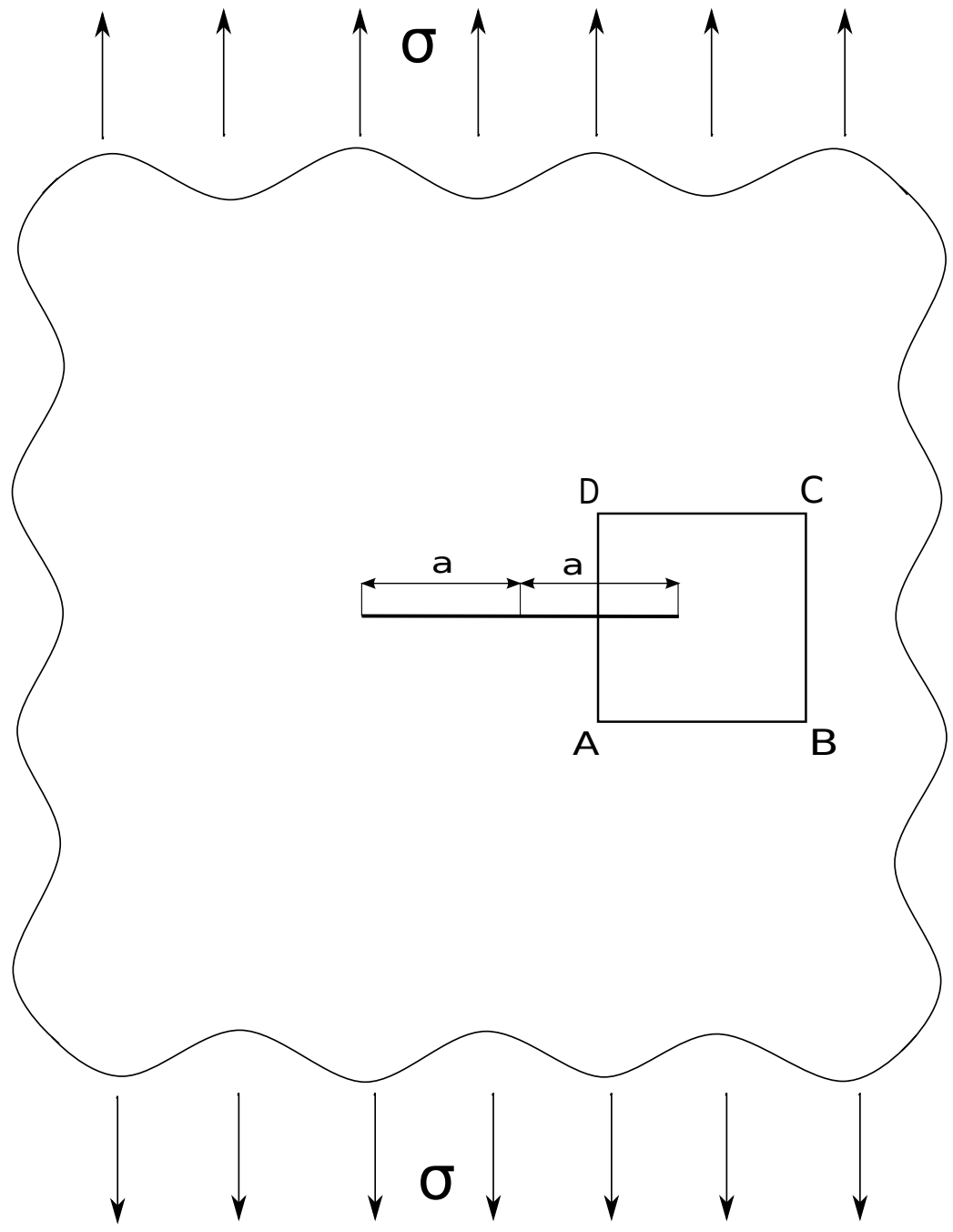

Figure 15: The infinite plate with a center crack under uniform tension and modeled geometry ABCD.

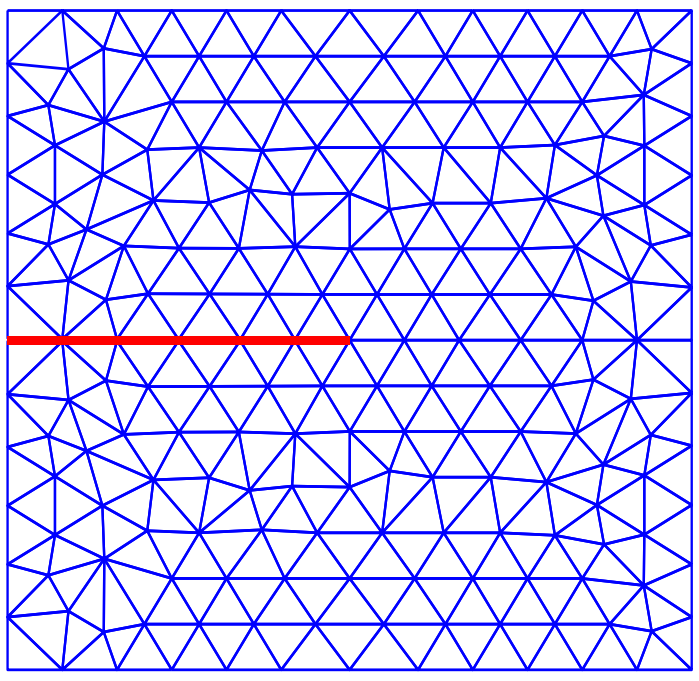

(a) T3 mesh (418 D.O.F)

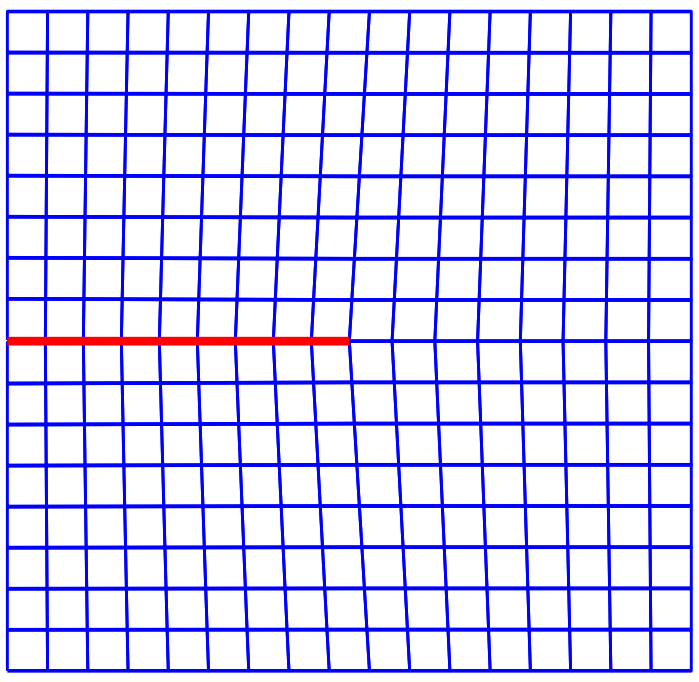

(b) Q4 mesh (630 D.O.F)

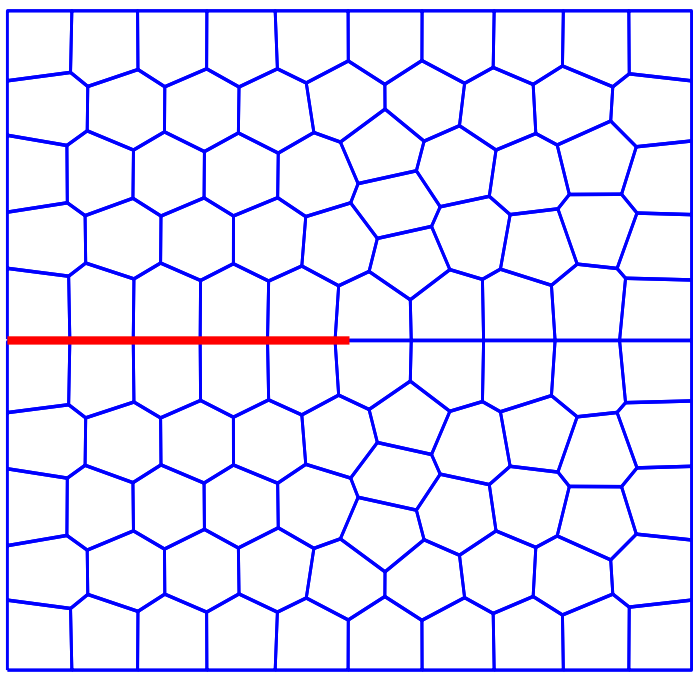

(c) Voronoi mesh (400 D.O.F)

Figure 16: Three coarse meshes for the asymptotic domain of the infinite plate with a center crack.

In analogy to the Timoshenko beam analysis, we investigate the convergence rate for the crack problem by doing the test on the asymptotic domain ABCD with the prescribed boundary conditions. Here, five meshes with descending element size $h$ for triangular, quadrilateral and polygonal elements are created as shown in Figure 16. Figure 17a shows the convergence rate of displacement norm, of which mathematical expression is given by equation (61) while Figure 17b illustrates the convergence rate of energy norm (62). Obviously, the slope of convergence rate in $L^2$-norm approaches to 1.0 and the energy norm is close to the theoretical value of 0.5 for three types of mesh. Both are close to the optimal convergence of discontinuous analysis without tip enrichment. As we can see from Figure 18, furthermore, when the analytical solution is replaced by the superconvergence as described in equation (64), the convergence rate still yields the same rate of 0.5.

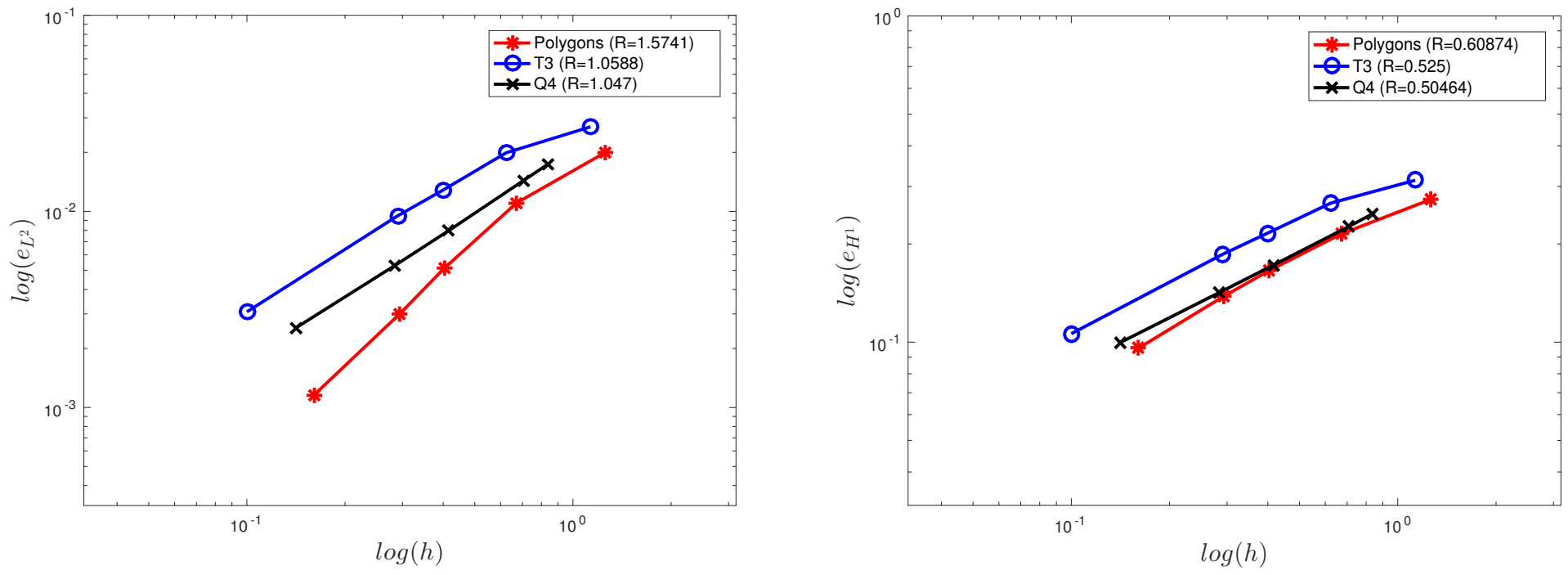

(a) (b)

Figure 17: Convergence rate of $L^2$-norm of error in displacement (a) and energy error norm (b) in the asymptotic crack domain

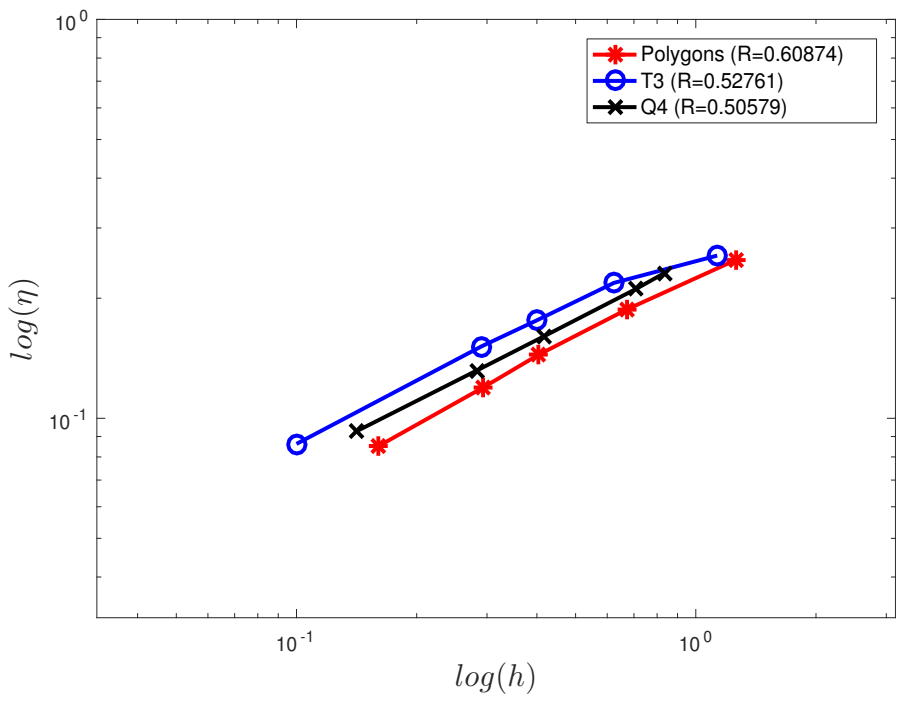

Figure 18: Convergence rate of energy error norm with assembled recovery stress in the asymptotic crack domain.

With the promising results of the recovery solution, an attempt to perform the mesh refinement is made. In this test we apply NVB, polyTree and midPoint to different meshes. As shown in Figure 19, 20, 21, because the singularity occurs at the crack tip and the high stress predominantly concentrates around it, the meshes are mainly refined in the vicinity of the crack tip. The condition to stop the refinement task in the test is when the total error reaches the critical value of less than 6% . As illustrated, VEM can circumvent the violated inter-element compatibility occurring in polyTree and midPoint refinements.

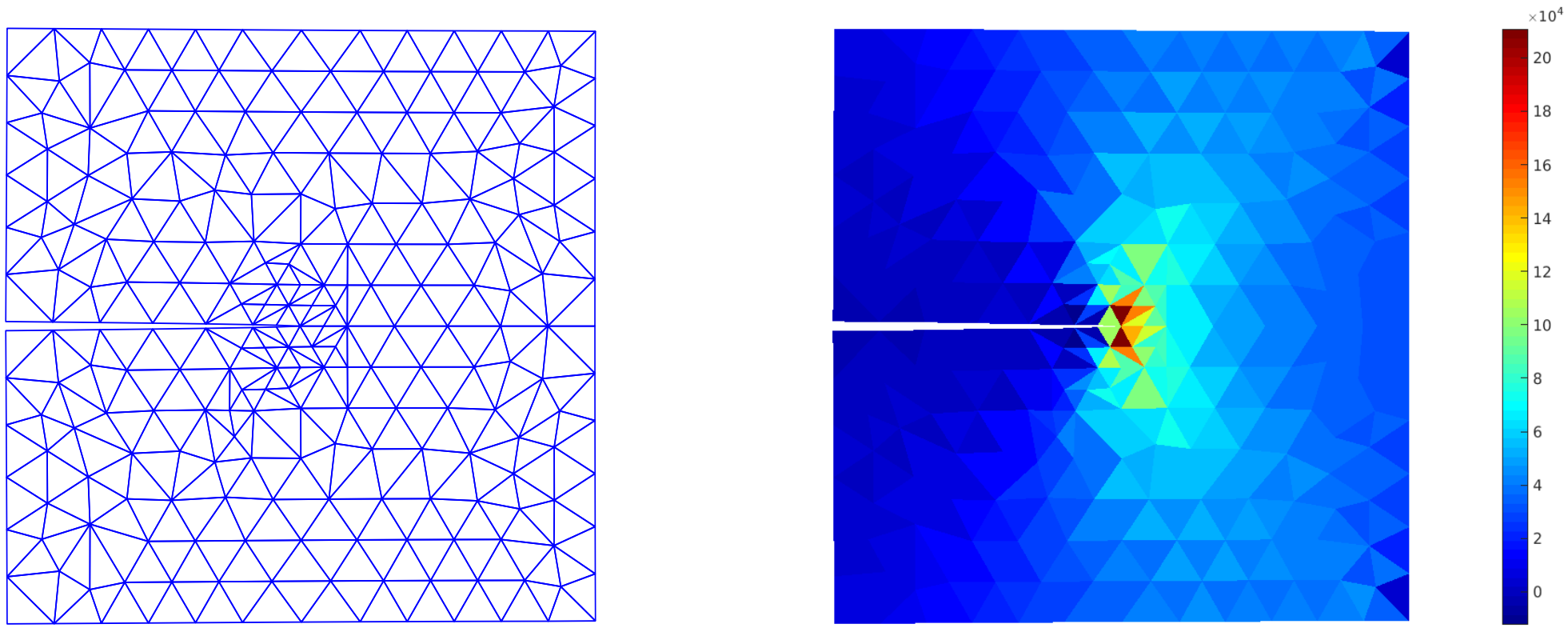

$\eta = 18.0533\%$ $\sigma_{yy}$ of the relevant mesh

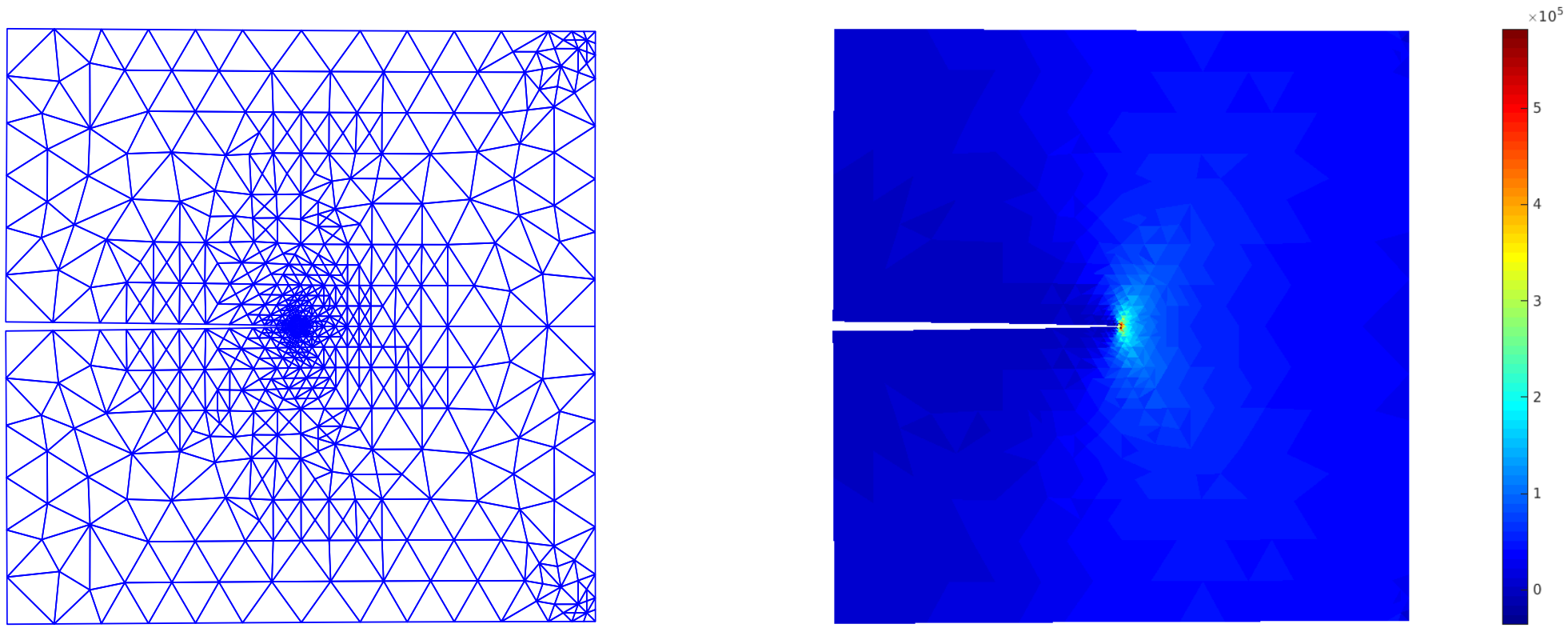

$\eta = 11.4597\%$ $\sigma_{yy}$ of the relevant mesh

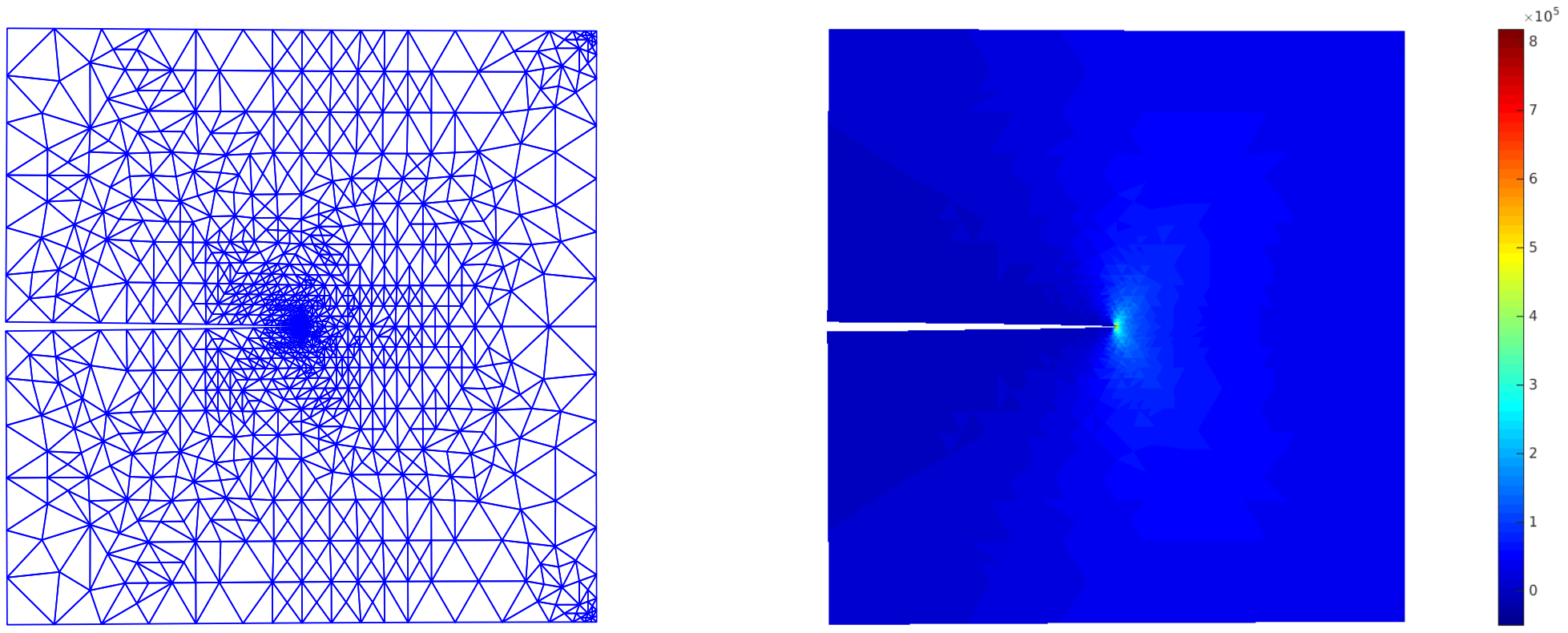

$\eta = 8.8262\%$ $\sigma_{yy}$ of the relevant mesh

Figure 19: Local adaptivity using NVB algorithm for the asymptotic domain.

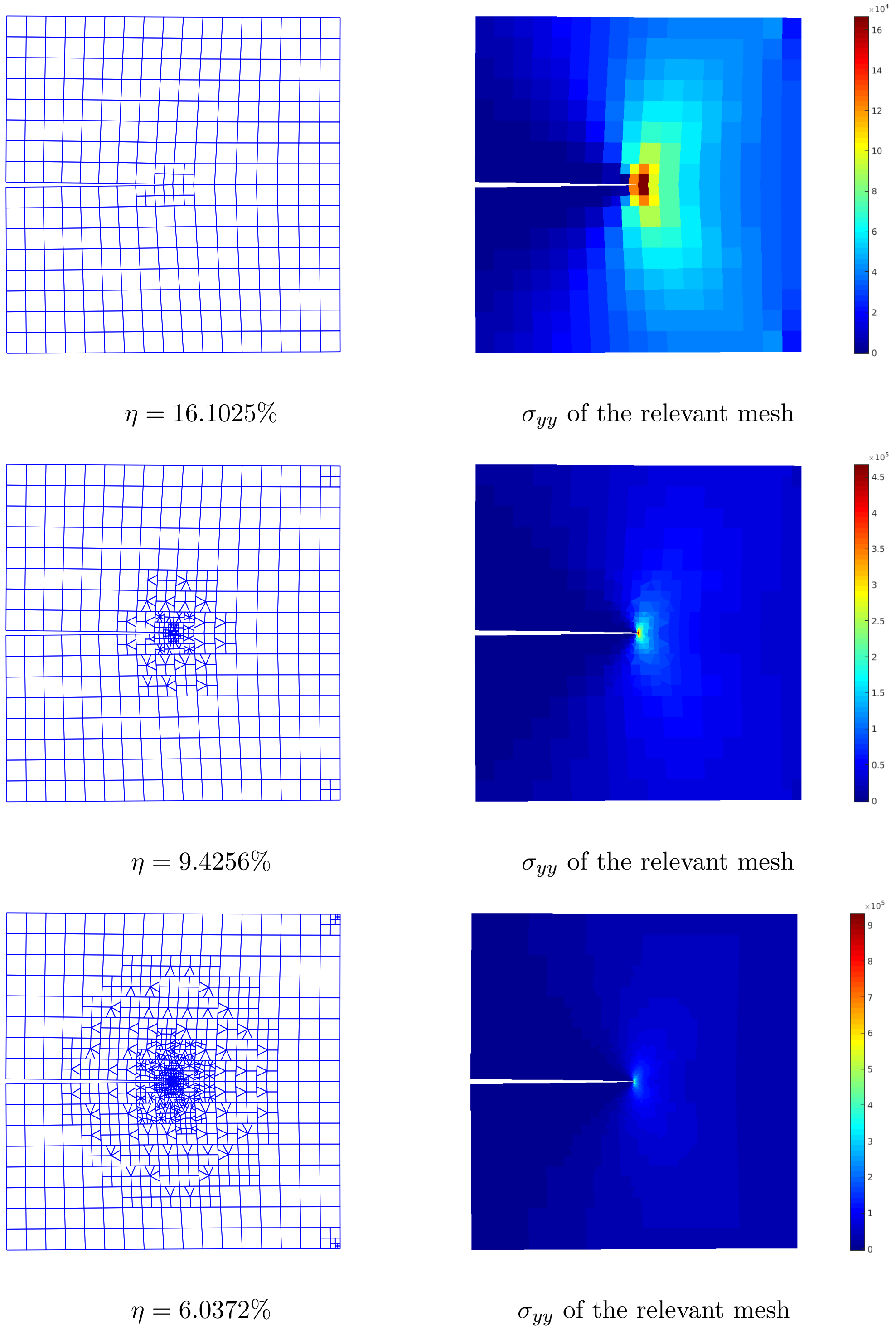

Figure 20: Local adaptivity using midPoint for the asymptotic domain.

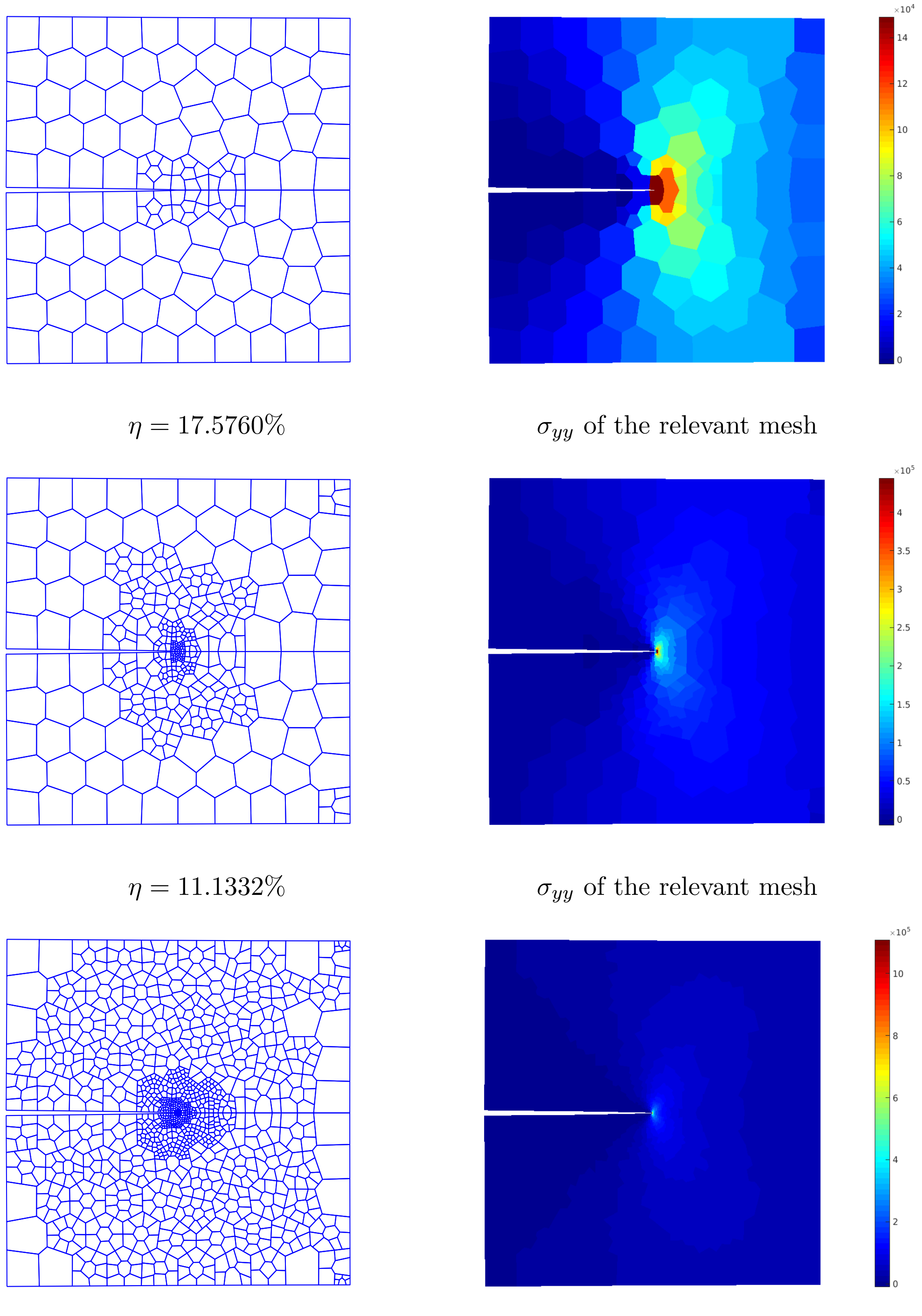

Figure 21: Local adaptivity using polyTree for the asymptotic domain.

### *4.3 Slanted crack in a biaxial stress field*

Next, we consider a slanted crack with the length of $2a$ in an infinite domain subjected to a biaxial stress field. The angle between the crack and the $x$ axis is defined by $\beta$ as shown in Figure 22. A two-dimensional domain configuration is created to model this problem with a half crack length of $a = 0.5\,\text{in}$, tension stress of $\sigma = 2000\,\text{psi}$. The width of our domain configuration is $W = 10\,\text{in}$ as shown in Figure 22b. Plane strain state is assumed with Young's modulus of $E = 3\times 10^7$ psi, Poisson's ratio of $\nu = 0.25$. In this example, the analytical solution of stress intensity factors are available in [51] and given as

$$K_I = \sigma\sqrt{\pi a}(\cos^2\beta + \alpha\sin^2\beta), \tag{70}$$

$$K_{II} = \sigma\sqrt{\pi a}(1-\alpha)\sin\beta\cos\beta. \tag{71}$$

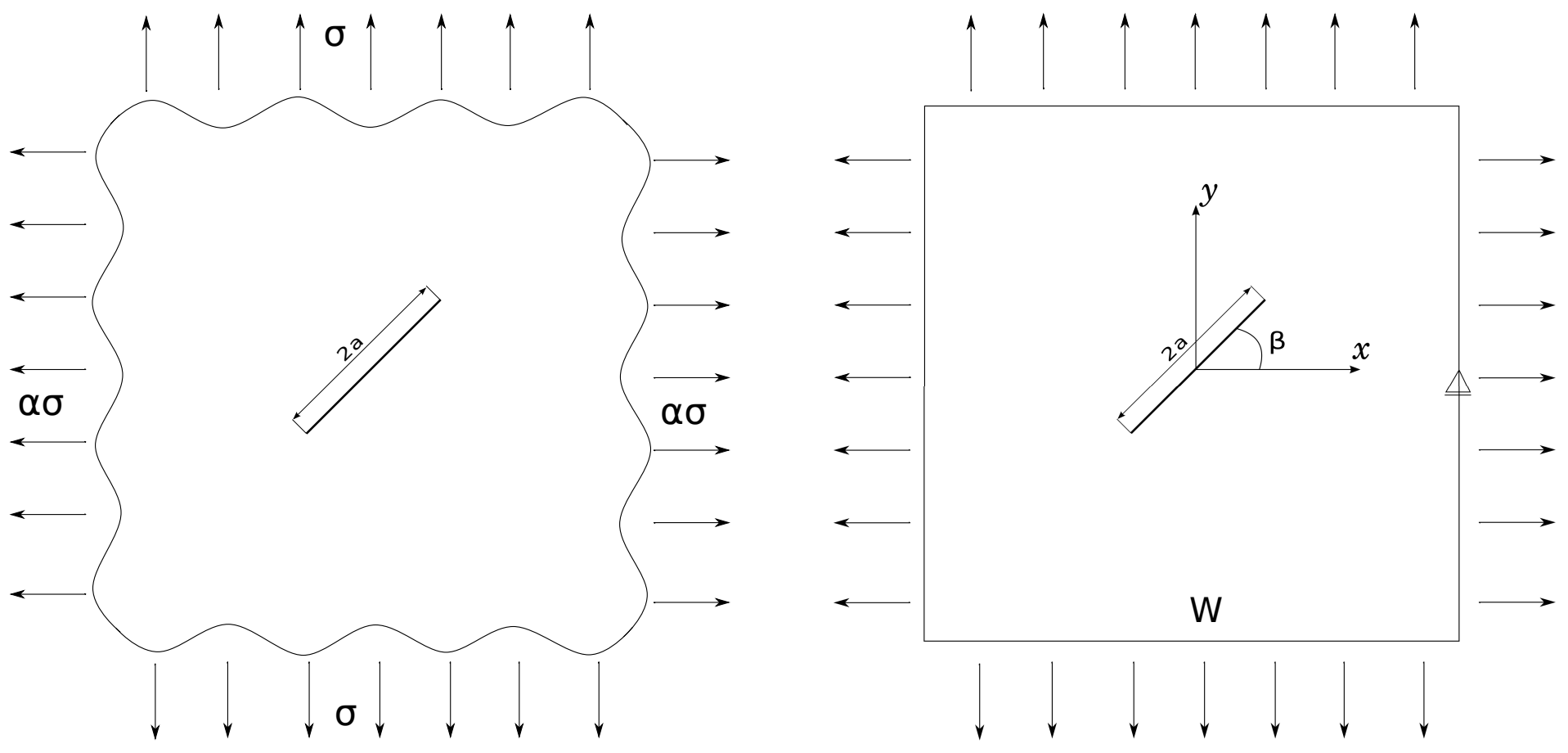

(a) (b)

Figure 22: Slanted crack in a biaxial stress field.

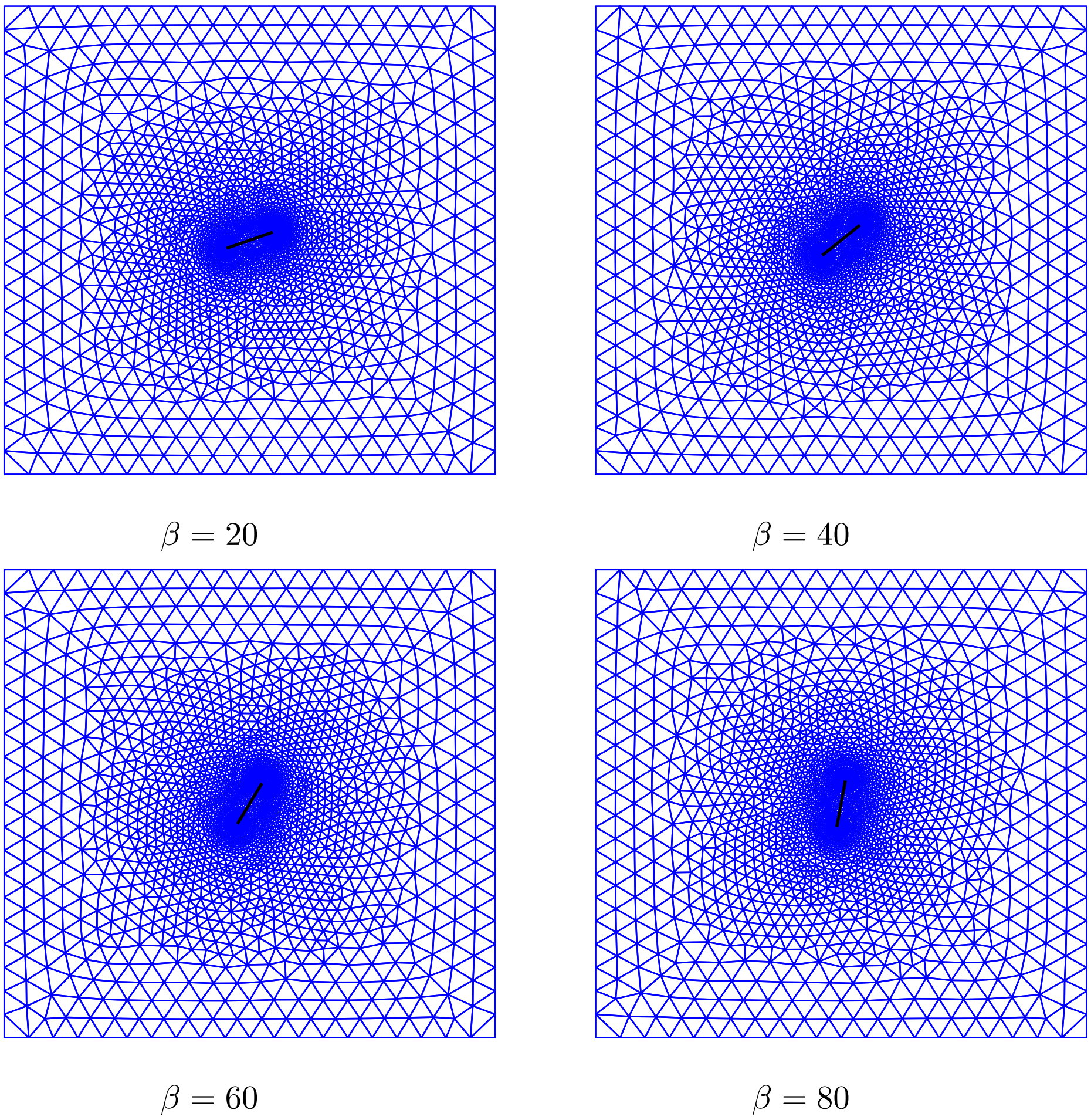

Figure 23: Triangular meshes for different angles in slanted crack problem.

We examine the case in which $\alpha$ is set to zero, so the problem becomes the slanted crack in different angles from 0 to 90 degree under the vertical pulling test [52]. Table 1 shows the comparison between our numerical results of SIFs calculated based on triangular discretization and analytical solutions ($K_I^{ex}$, $K_{II}^{ex}$) in addition to the normalized SIFs. We also plot the numerical results of SIF in mode $I$ and mode $II$ by discrete points and the theoretical curve in Figure 25. The result shows a good agreement between the exact solution and the numerical one for the entire range of $\beta$. Furthermore, we investigate which direction the crack is propagating when the crack opens in three cases of $\alpha$ (0, 0.25, and 0.5). As shown in Figure 25, the numerical

| Crack angle $\beta$ | $K_I$ | $K_I^{ex}$ | $K_I/K_I^{ex}$ | $K_{II}$ | $K_{II}^{ex}$ | $K_{II}/K_{II}^{ex}$ |
|---|---|---|---|---|---|---|
| 0 | 2530.3184 | 2506.6282 | 1.0094 | 12.7906 | 0 | - |
| 10 | 2444.0189 | 2431.0441 | 1.0053 | 431.3651 | 428.6587 | 1.0063 |
| 20 | 2222.9372 | 2213.4084 | 1.0043 | 801.9698 | 805.6148 | 0.9954 |
| 30 | 1885.8464 | 1879.9712 | 1.0031 | 1083.4897 | 1085.4019 | 0.9982 |
| 40 | 1502.1929 | 1470.9498 | 1.0212 | 1198.5552 | 1234.2735 | 0.9710 |
| 50 | 1034.1677 | 1035.6784 | 0.9985 | 1231.3324 | 1234.2735 | 0.9976 |
| 60 | 632.2798 | 626.6570 | 1.0089 | 1086.2555 | 1085.4019 | 1.0007 |
| 70 | 302.8797 | 293.2198 | 1.0329 | 812.9874 | 805.6148 | 1.0091 |
| 80 | 67.6346 | 75.5840 | 0.8948 | 423.9832 | 428.6587 | 0.9890 |
| 90 | -1.1440 | 0 | - | 93.8169 | 0 | - |

Table 1: Numerical results of SIFs and normalized SIFs at different angles in the slanted crack problem.

results fit well with the analytical curves where $\alpha^*$ denotes the crack growth angle.

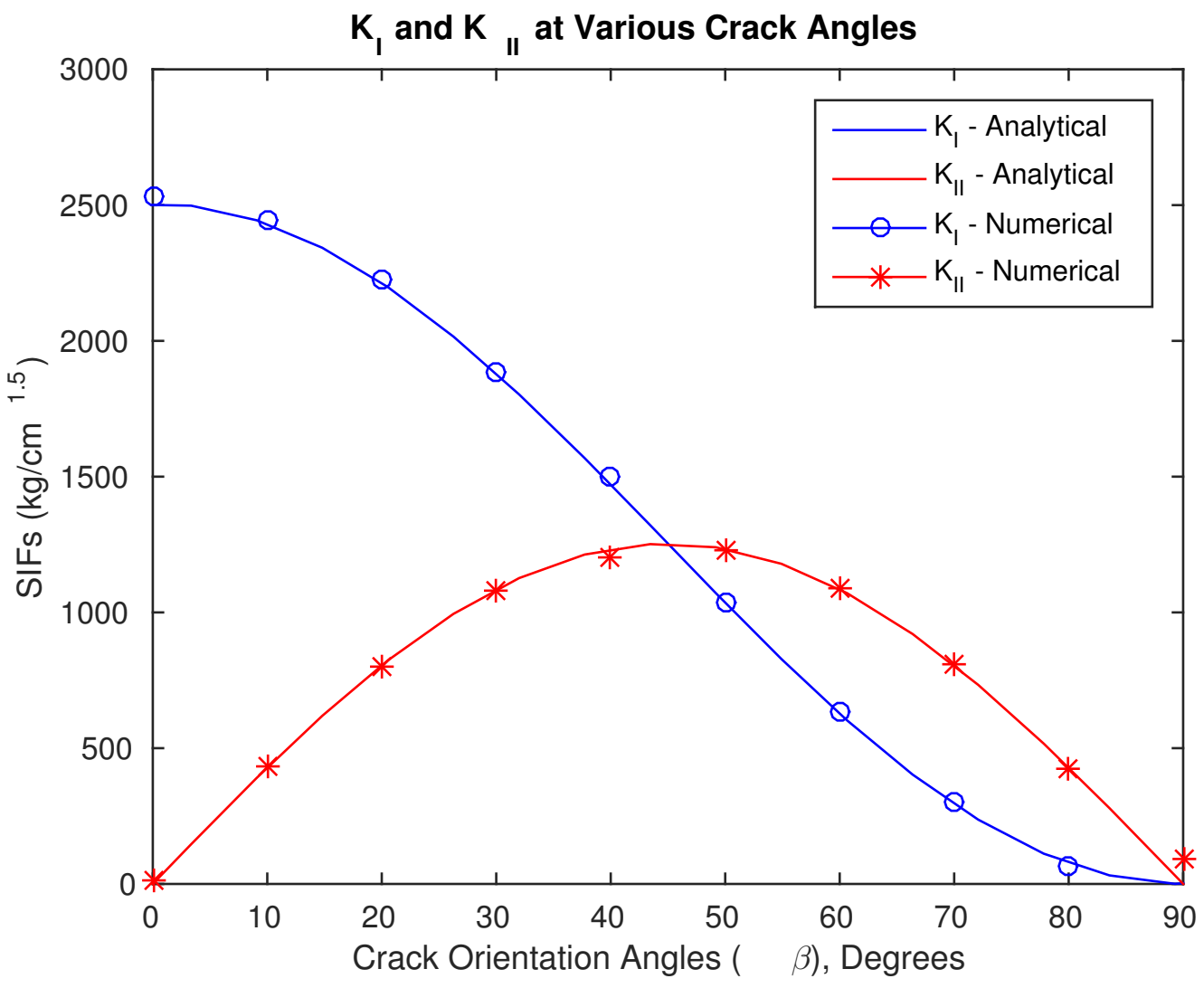

Figure 24: $K_I$ and $K_{II}$ at various crack angles.

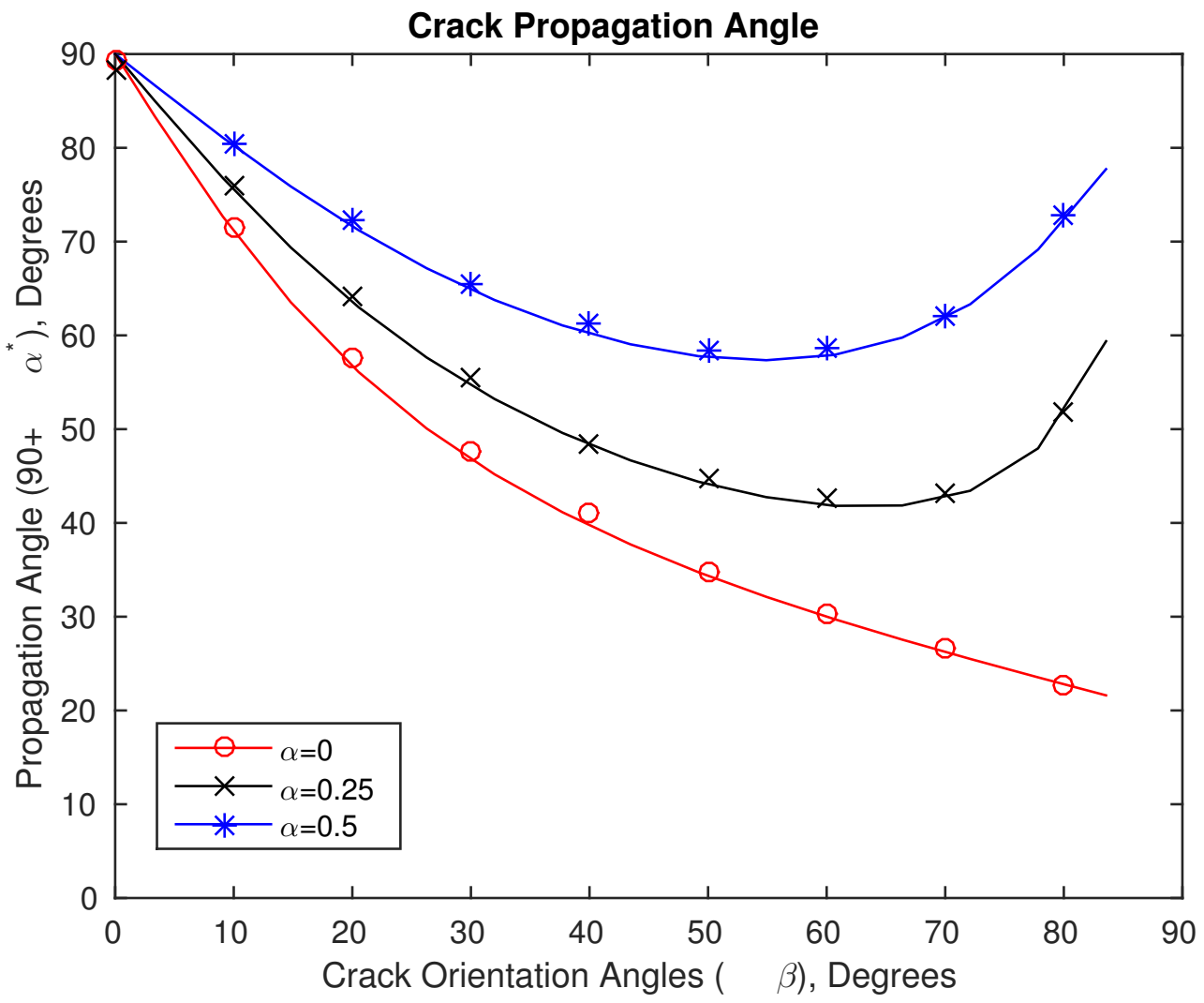

Figure 25: Crack propagation angle at initial opening crack in different $\alpha$.

### *4.4 Crack propagation of an edge-crack under mixed-mode loading*

Let us take the edge crack subjected to a shear load $\tau = 1\,\mathrm{N/mm^2}$ , schematically shown in Figure 26 as an example. A plate with the width 7 mm, the height 16 mm, contains an edge crack with the length of $a = 3.5$ mm. In this example, Young's modulus $E = 3 \times 10^7\,\mathrm{N/mm^2}$ and Poisson's ratio $\nu = 0.25$ are the material parameters. The plane stress state is assumed. The exact stress intensity factors can be found in the literature as follows

$$K_I = 34\,\mathrm{mm}^{-3/2},$$
$$K_{II} = 4.55\,\mathrm{mm}^{-3/2}.$$

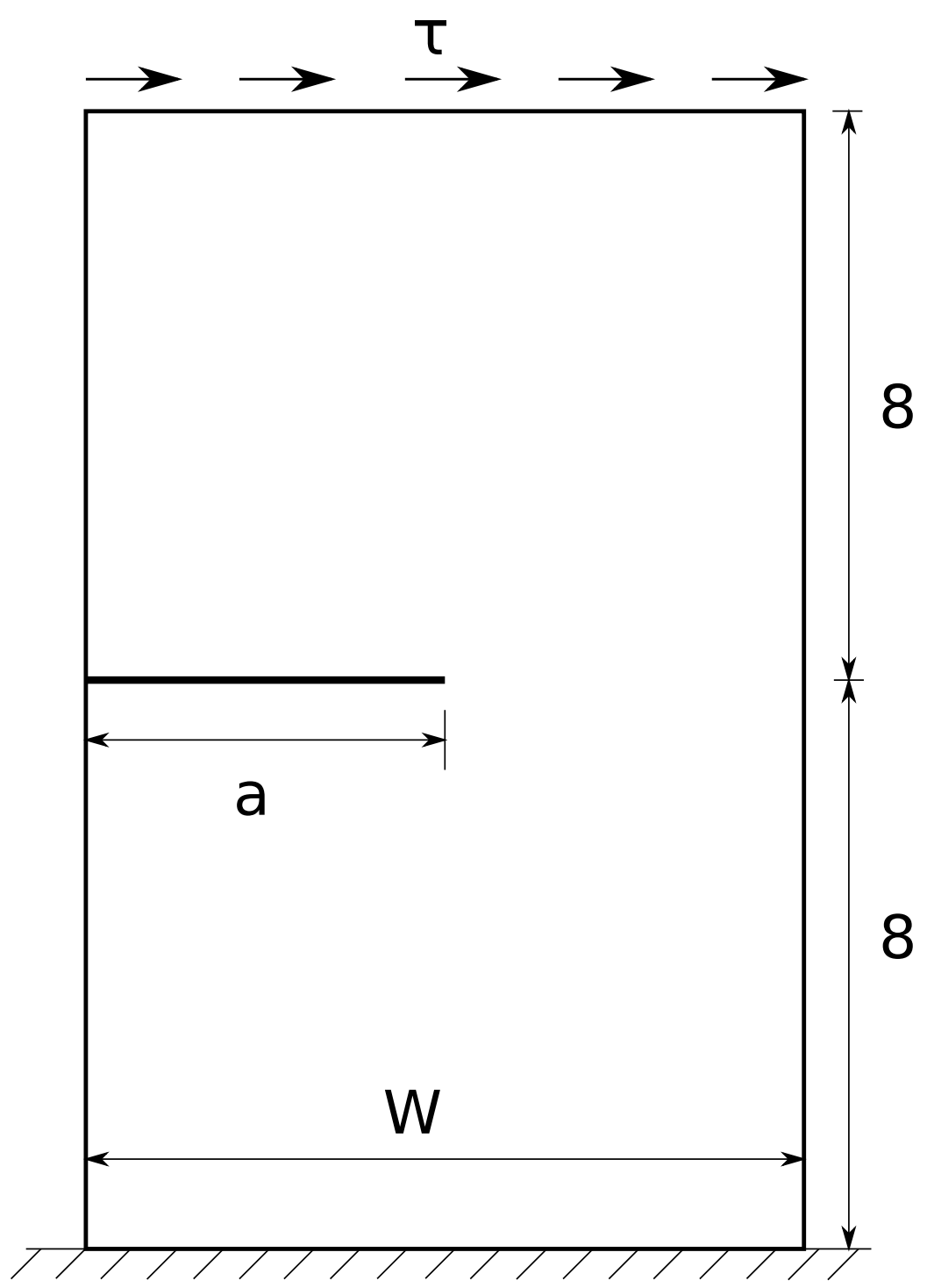

Figure 26: Shear edge crack.

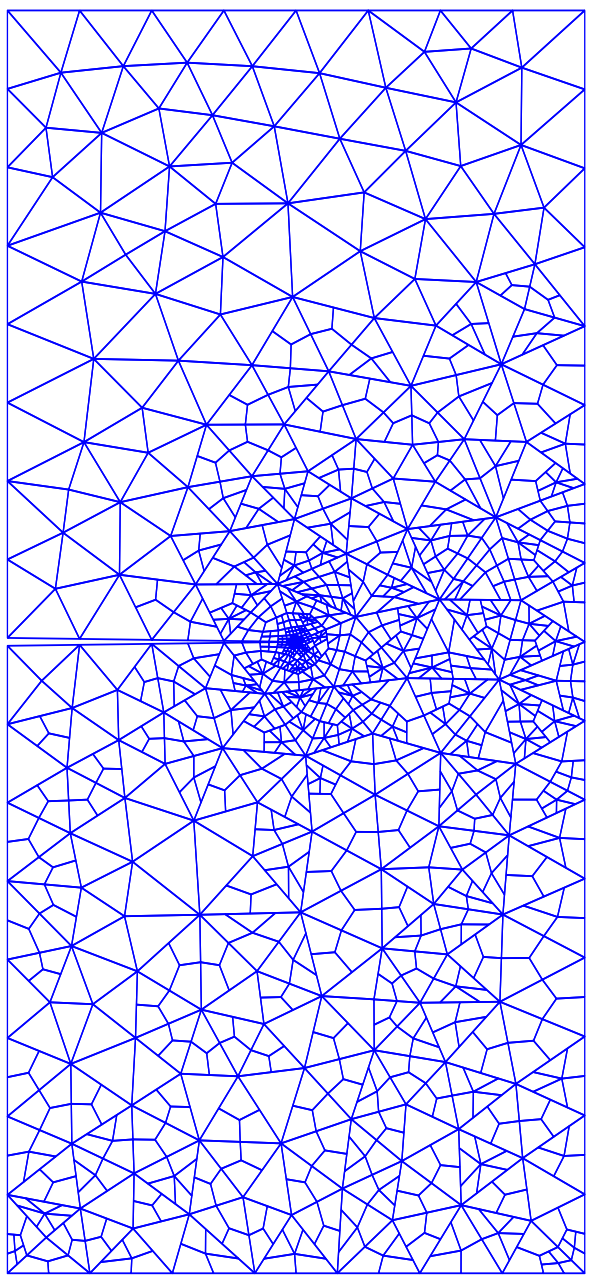

Step 1

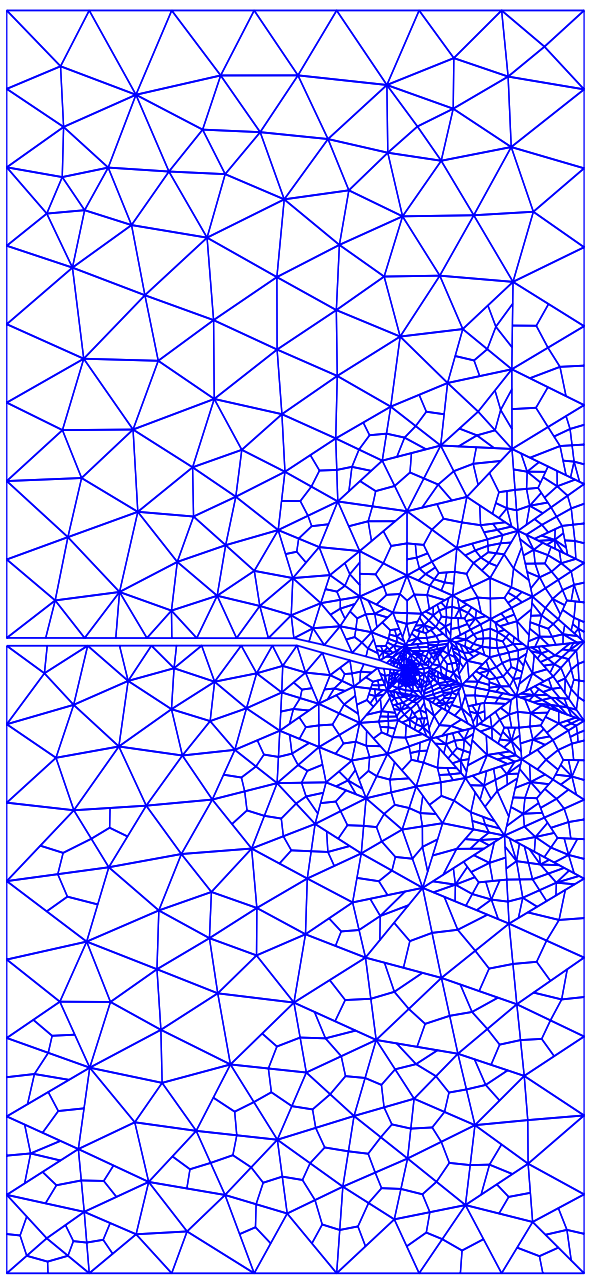

Step 5

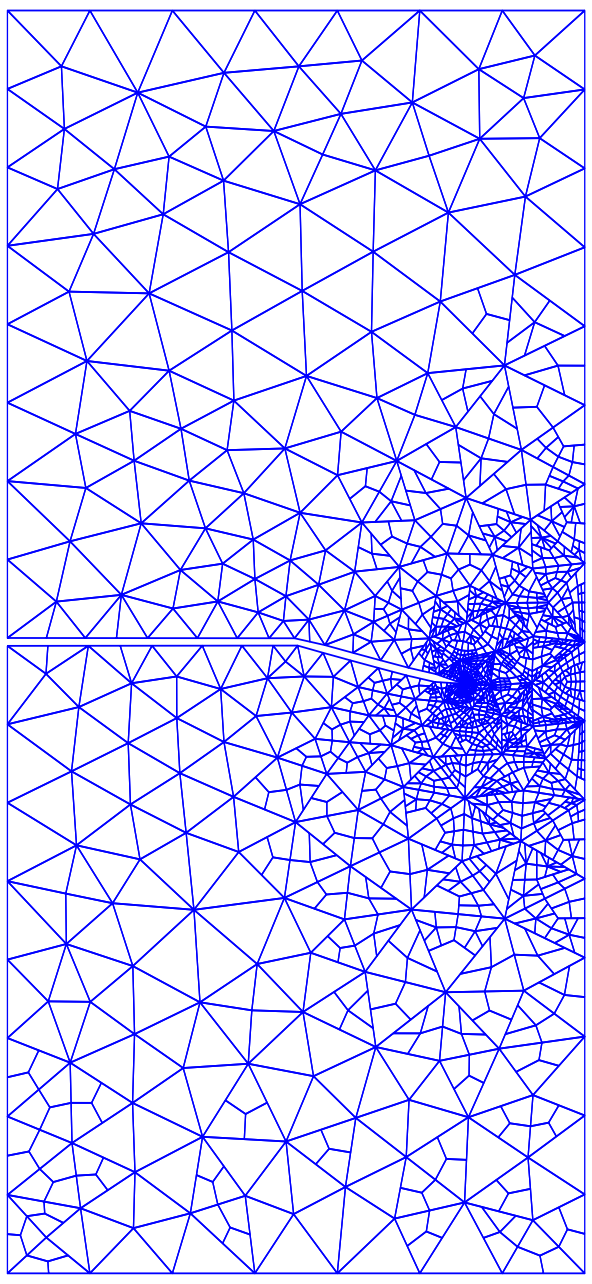

Step 7

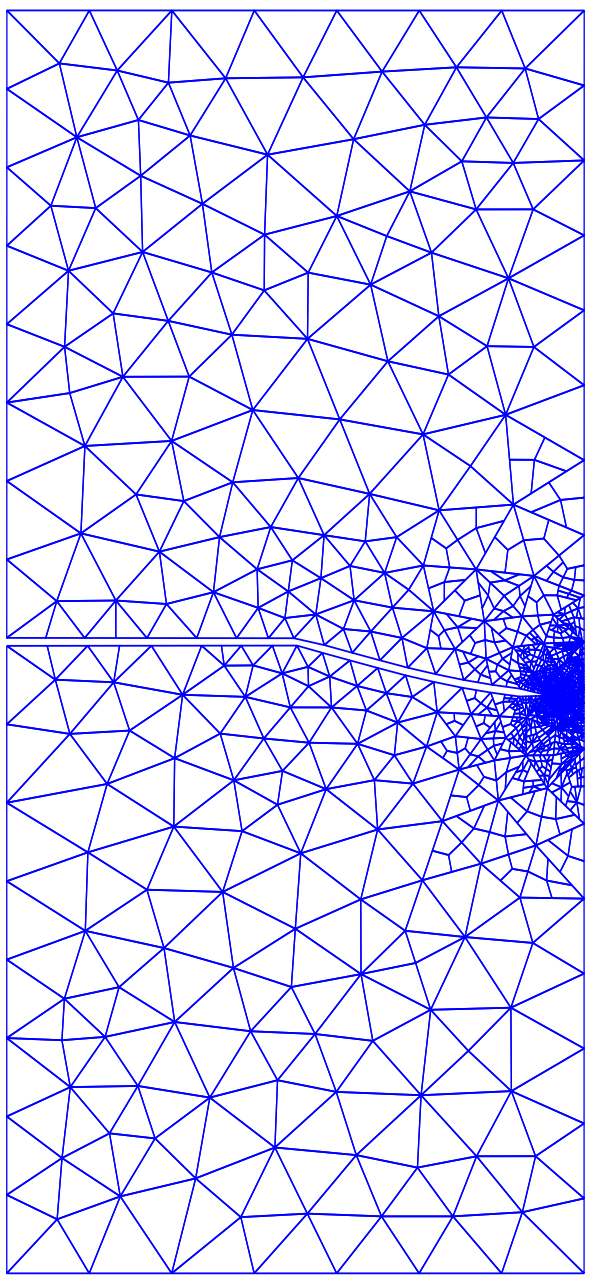

Step 9

Figure 27: A typical adaptive mesh refinement for edge crack growth.

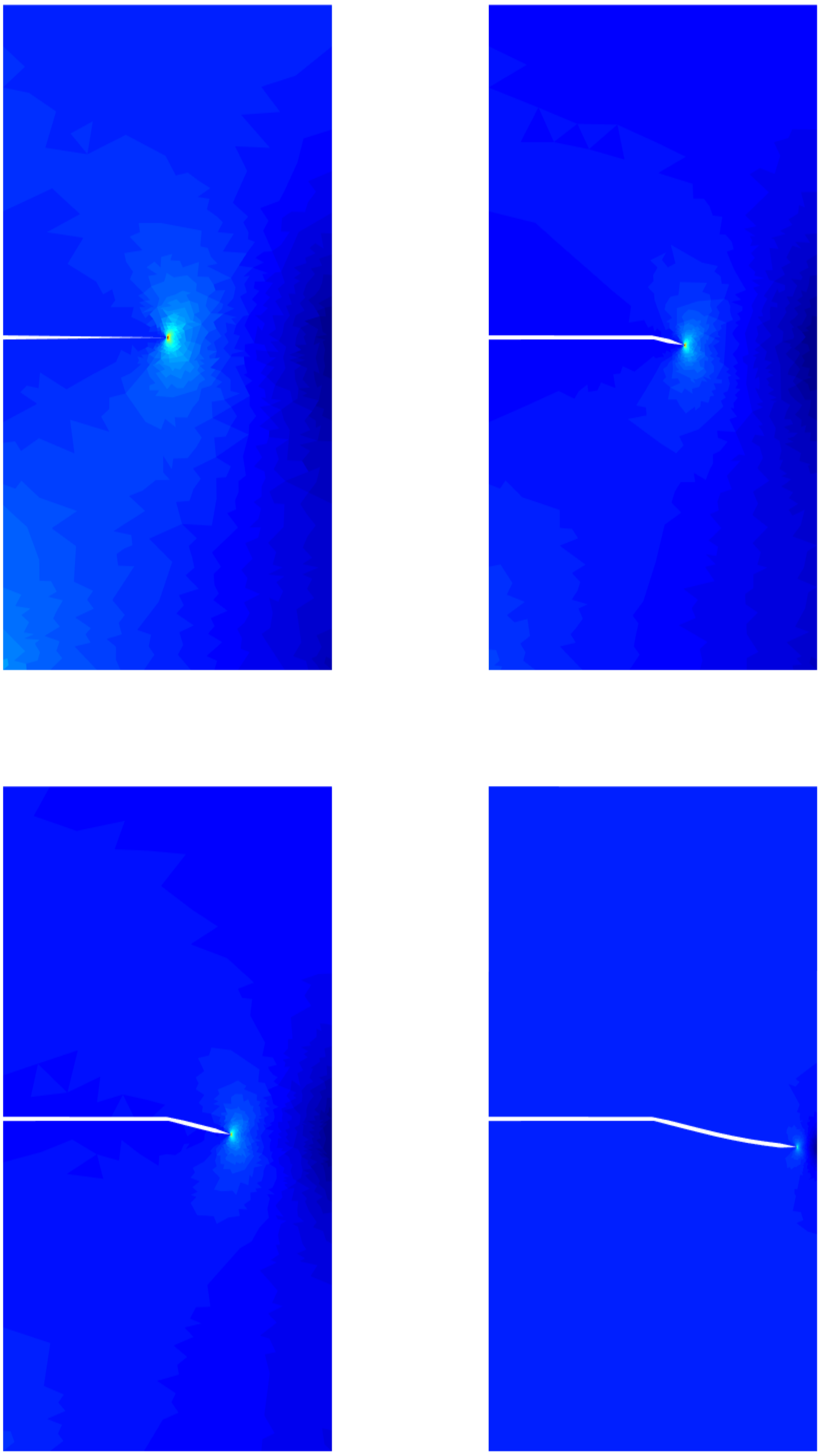

Figure 28: The development of the crack in different steps.

Figure 27 illustrates several meshes in the adaptive mesh refinement loop with respect to the total error $\eta$ calculated in equation (64). The refinement is repeated until the error is less than the critical value, here we take 10% in our numerical experiment. The first coarse mesh is initiated with triangular elements. Then, the next adaptive meshes are refined based on midPoint refinement. Obviously, VEM allows us to flexibly deal with the arbitrary mesh even with the presence of hanging nodes. The refinement mainly occurs in the vicinity of crack tip. Also, elements at the bottom and right side of the plate are subdivided into small elements. These areas are under high stresses, especially in the crack tip. With the final mesh, we are able to obtain SIFs and the pertinent propagation angle. From this step a new mesh is generated based on a new geometry updated from the crack growth. The crack propagation is schematically shown in Figure 28. It shows that in this problem the crack tends to slightly go down. Compared with other numerical methods [53], the crack propagates through a correct path represented in totally 10 steps.

### *4.5 Crack propagation of a panel with rivet holes*

The crack propagation of a polymethyl methacrylate (PMMA) beam with three initial holes is simulated in this example as shown in Figure (29). The beam has length $L = 20$, width $W = 8$. A concentrated load is applied at the center of the top edge of the PMMA beam. The parameters are given as: Young's modulus $E = 4 \times 10^5$ psi, Poisson's ratio $\nu = 0.3$, load $P = 1$ lb. The initial crack length is given as $a = 1$. Plane strain state is assumed. The experimental and numerical tests for the PMMA have been observed in [53, 54, 55].

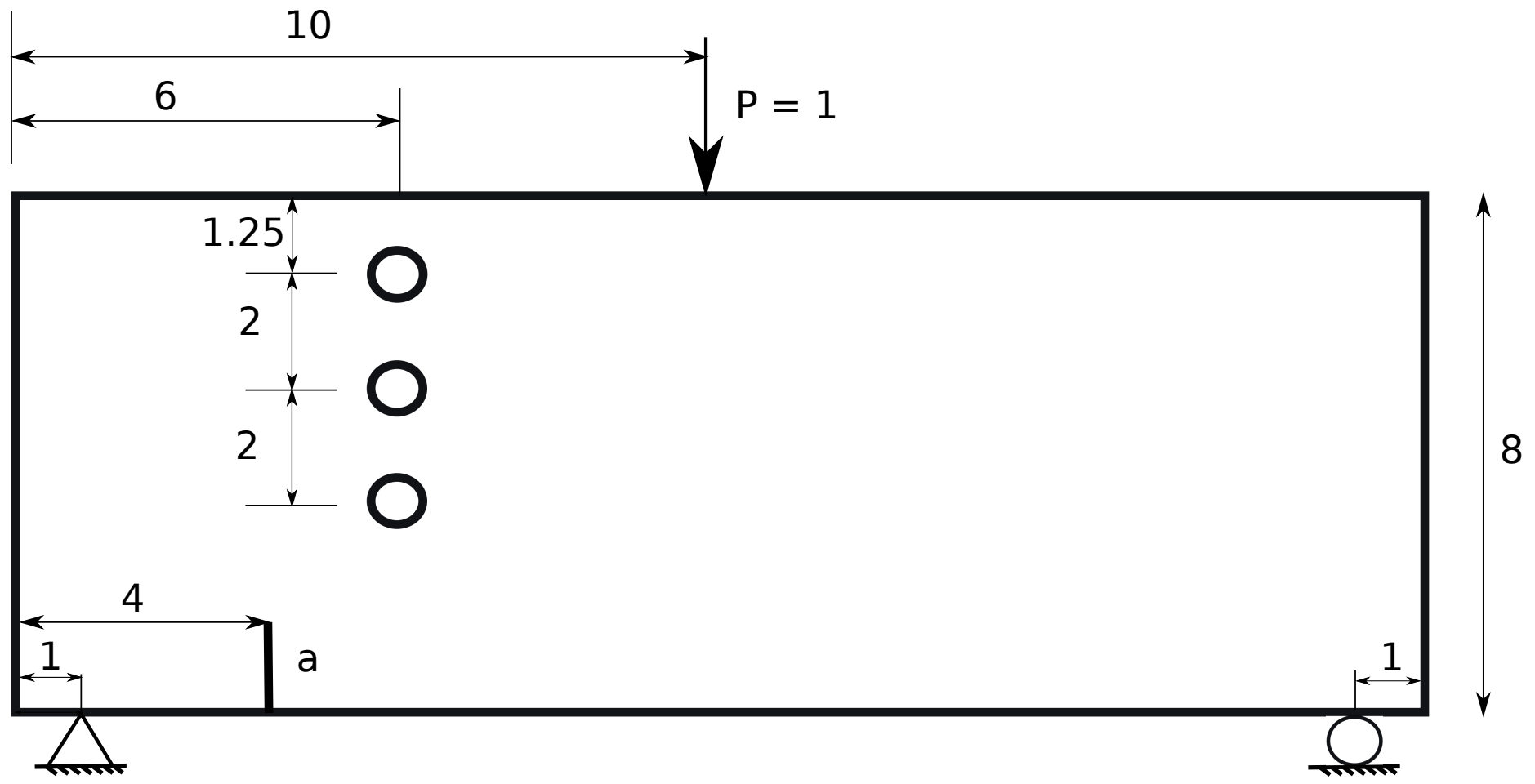

Figure 29: PMMA beam with three holes subjected to a concentrated load.

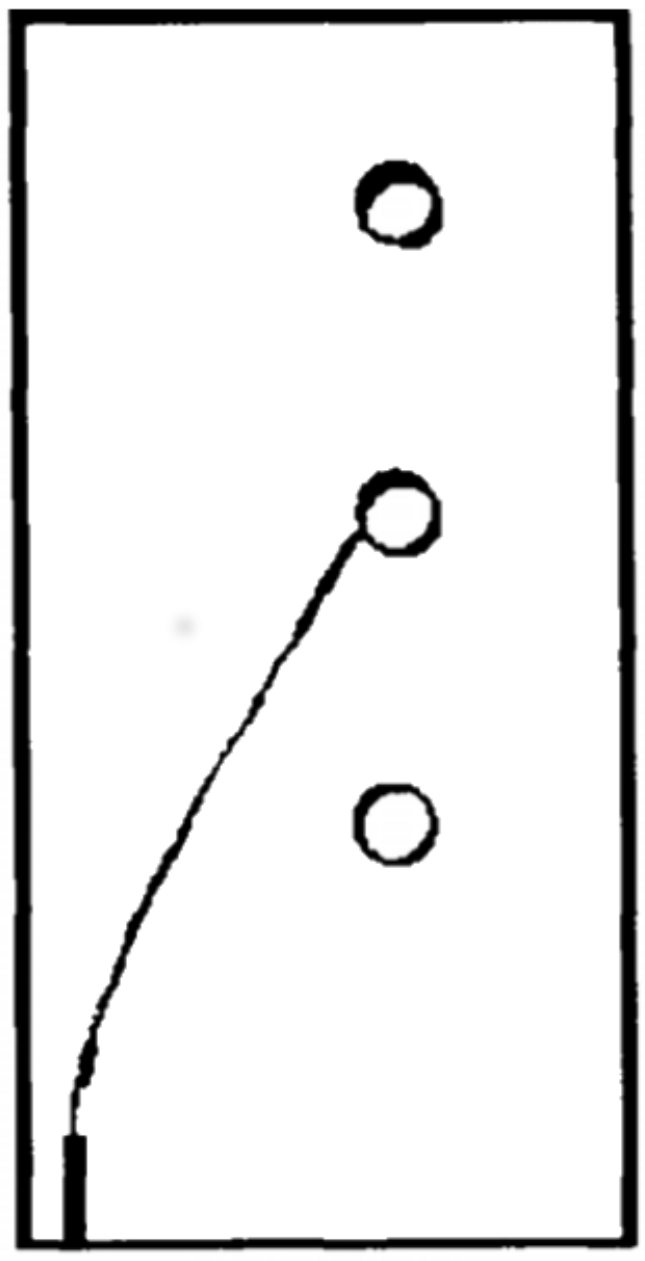

Figure 30: Experimental crack growth trajectory for PMMA test [54].

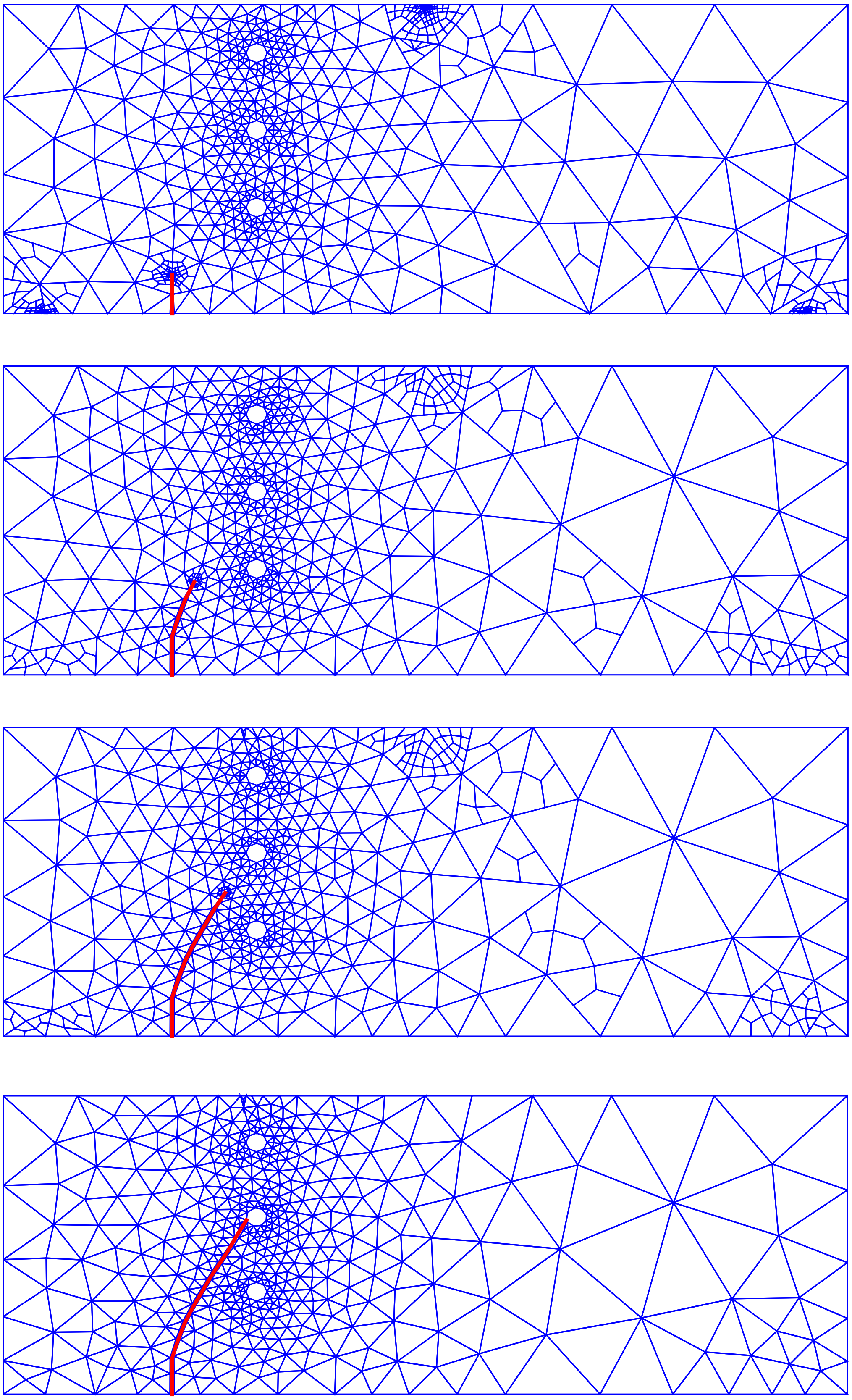

Figure 31: Crack growth trajectory for the PMMA beam.

In this analysis the triangular coarse mesh is first initiated and then the adaptive scheme is performed by using the midPoint refinement. As we can see in Figure 31, the refined meshes contain mostly triangular, quadrilateral, or polygonal elements with hanging nodes. The final crack path is found after 11 steps and it cuts through the middle hole. Experimental observation shown in Figure 30 is compared to our predicted crack and the path shows the excellent agreement.

## 5 Conclusion

In this paper, we have developed a formulation for the VEM to model 2D linear elastic fracture problems. Regarding the local stiffness matrix in VEM, we use a stability term which is stable in terms of both isotropic scaling and ratio. The main advantage of the present scheme is that all calculation and derivation are performed based on the polygonal elements. With the aid of VEM, adaptive meshes allow the presence of hanging nodes in computation. The SIFs are calculated through a set of arbitrary polygonal elements surrounding the vicinity of the crack tip. The stress concentration at the crack tip, the displacement discontinuity, and crack propagation are captured by the adaptive mesh refinement. The adaptive strategies such as longest vertex bisection, middle point and polyTree for triangular, quadrilateral and polygonal discretization, respectively, are investigated with the proposed posteriori error estimates. The numerical examples show the optimal convergence for both the continuous and discontinuous problems. They also show that SIFs and propagation angle in the numerical results agree well with the theoretical and experimental results.

## 6 Acknowledgments

The first and second authors owe the gratitude to the sponsorship from Sofja Kovalevskaja Programme of Alexander von Humboldt Foundation.